\RequirePackage{fix-cm}

\documentclass[smallextended]{svjour3} 

\smartqed 
\usepackage{xspace}
\usepackage{ifthen}
\usepackage{alltt}
\usepackage{amsmath}    
\usepackage{amsfonts}
\usepackage{amssymb}
\usepackage{latexsym}
\usepackage{balance}
\usepackage{booktabs}
\usepackage{enumerate}
\usepackage[T1]{fontenc} %%%key to get copy and paste for the code!
\usepackage{graphicx}
\usepackage[pdftex,colorlinks=true,pdfstartview=FitV,linkcolor=black,citecolor=black,urlcolor=black,bookmarks=false]{hyperref}
\usepackage[scaled=0.85]{helvet}
\usepackage[utf8]{inputenc} %%% to support copy and paste with accents for frnehc stuff
\usepackage{lscape}
\usepackage{listings} %for Smalltalk code
\usepackage{lipsum}
\usepackage{mathtools}
\usepackage{multicol} % for multicolumns in equation figures
\usepackage{needspace}
\usepackage{stmaryrd}
\usepackage{times}
\usepackage[most]{tcolorbox}
\usepackage[normalem]{ulem} % for \sout
\usepackage{textcomp}
\usepackage{verbatim}
\usepackage[modulo]{lineno} % for line numbers
\usepackage{paralist} %to list extends of previous work
\usepackage[numbers]{natbib} % for numeric citations
\usepackage{lstautogobble}

\usepackage{enumitem} %for description in enumeration

\definecolor{source}{gray}{0.95}
\newcommand{\code}[1]{\texttt{#1}}

\usepackage{ifthen}
\usepackage{xcolor,colortbl}

\newboolean{showedits}
\setboolean{showedits}{true} % toggle to show or hide edits
\ifthenelse{\boolean{showedits}}
{
	\newcommand{\del}[1]{\textcolor{red}{\sout{#1}}} % please delete
	\newcommand{\nbe}[3]{
		{\colorbox{#3}{\bfseries\sffamily\scriptsize\textcolor{white}{#1}}}
		{\textcolor{#3}{\sf\small$\blacktriangleright$\textit{#2}$\blacktriangleleft$}}}
}{
	\newcommand{\del}[1]{} % please delete
	
	\newcommand{\nbe}[3]{}
}

\newboolean{showcomments}
\setboolean{showcomments}{true} 
\setboolean{showcomments}{false} 
\newcommand{\id}[1]{$-$Id: scgPaper.tex 32478 2010-04-29 09:11:32Z oscar $-$}

\ifthenelse{\boolean{showcomments}}
 {
 	\newcommand{\nbc}[3]{
 		{\colorbox{#3}{\bfseries\sffamily\scriptsize\textcolor{white}{#1}}}
		{\textcolor{#3}{\sf\small$\blacktriangleright$\textit{#2}$\blacktriangleleft$}}}
	
 }{
 	\newcommand{\nbc}[3]{}
 	
 }

\usepackage[most]{tcolorbox}
\ifthenelse{\boolean{showedits}}
{
  \newtcolorbox{inserted}{%
       title=Inserted text:,
       colframe=blue,colback=blue!5!white,
       breakable,
       leftrule=0mm, 
       bottomrule=0mm,
       rightrule=0mm,
       toprule=0mm,
       arc=0mm, outer arc=0mm,
       oversize
  }
  \newtcolorbox{deleted}{%
       title=Deleted text:,
       colframe=red,colback=red!5!white,
       breakable,
       leftrule=0mm, 
       bottomrule=0mm,
       rightrule=0mm,
       toprule=0mm,
       arc=0mm, outer arc=0mm,
       oversize
  }
  \newtcolorbox{refactored}{%
       title=Rewritten text:,
       colframe=blue,colback=red!5!white,
       breakable,
       leftrule=0mm, 
       bottomrule=0mm,
       rightrule=0mm,
       toprule=0mm,
       arc=0mm, outer arc=0mm,
       oversize
  }
}{

}

\newcommand{\commented}[1]{}
\newcommand{\eg}{\emph{e.g.,}\xspace}
\newcommand{\ie}{\emph{i.e.,}\xspace}
\newcommand{\etal}{\emph{et al.}\xspace}
\newcommand{\etc}{\emph{etc.}\xspace}
\makeatletter                  
\def\mdseries@tt{m}      
\makeatother                   
\journalname{Empirical Software Engineering}

\begin{document}

\title{Security Code Smells in {Android} {ICC}}

\author{Pascal Gadient \and Mohammad Ghafari \and \\ Patrick Frischknecht \and Oscar Nierstrasz}

\institute{Pascal Gadient, Mohammad Ghafari, Patrick Frischknecht, Oscar Nierstrasz \at
Software Composition Group, University of Bern, Switzerland\\
\email{\{gadient, ghafari, oscar\}@inf.unibe.ch}
}

\date{Received: date / Accepted: date}

\maketitle

% ============================================================
\newcommand{\smell}[1]{\textbf{#1}.}

\newcommand{\Issue}{\emph{Issue:}\xspace}
\newcommand{\Consequences}{\emph{Consequently},\xspace}
\newcommand{\consequences}{\emph{consequently},\xspace}
\newcommand{\Symptoms}{\emph{Symptom:}\xspace}
\newcommand{\Mitigation}{\emph{Mitigation:}\xspace}

\newcommand{\Detection}{\emph{Detection:}\xspace}
\newcommand{\Limitations}{\emph{Limitation:}\xspace}
% ============================================================

\begin{abstract}

Android Inter-Component Communication (ICC) is complex, largely unconstrained, and hard for developers to understand.
As a consequence, ICC is a common source of security vulnerability in Android apps.
To promote secure programming practices, we have reviewed related research, and identified avoidable ICC vulnerabilities in Android-run devices and the security code smells that indicate their presence.
We explain the vulnerabilities and their corresponding smells, and we discuss how they can be eliminated or mitigated during development.
We present a lightweight static analysis tool on top of Android Lint that analyzes the code under development and provides just-in-time feedback within the IDE about the presence of such smells in the code.
Moreover, with the help of this tool we study the prevalence of security code smells in more than 700 open-source apps, and manually inspect around 15\% of the apps to assess the extent to which identifying such smells uncovers ICC security vulnerabilities.


\end{abstract}

% ==================================================
\section{Introduction} \label{sec:introduction}

Smartphones and tablets provide powerful features once offered only by computers.
However, the risk of security vulnerabilities on these devices is tremendous: smartphones are increasingly used for security-sensitive services like e-commerce, e-banking, and personal healthcare, which make these multi-purpose devices an irresistible target of attack for criminals.

A recent survey in the Stack Overflow website shows that about 65\% of mobile developers work with Android~\cite{StackOverflow:2017}.
%\footnote{\url{http://insights.stackoverflow.com/survey/2017}}
This platform has captured over 80\% of the smartphone market,
%\footnote{\url{http://www.gartner.com}}
and just its official app store contains more than 2.8 million apps.
As a result, a security mistake in an in-house app may jeopardize the security and privacy of billions of users.

The security of smartphones has been studied from various perspectives such as the device manufacturer~\cite{Wu:2013}, its platform~\cite{Xu:2016}, and end users~\cite{Jones:2015}.
Numerous security APIs, protocols, guidelines, and tools have been proposed.
Nevertheless security concerns are often overridden by other concerns~\cite{Balebako:2014}.
Many developers undermine their significant role in providing security~\cite{Xie:2011}.
As a result, security issues in mobile apps continue to proliferate unabated~\cite{CVE:2018}.
%\footnote{\url{http://www.cvedetails.com}}

Given this situation, in previous work we identified 28 security code smells, \ie symptoms in the code that signal potential security vulnerabilities~\cite{Ghafari:2017}.
We studied the prevalence of ten such smells, and realized that despite the diversity of apps in popularity, size, and release date, the majority suffer from at least three different security smells, and such smells are in fact good indicators of actual security vulnerabilities.

To promote the adoption of secure programming practices, we build on our previous work, and identify security smells related to Android Inter-Component Communication (ICC).
Android ICC is complex, largely unconstrained, and hard for developers to understand, and it is consequently a common source of security vulnerabilities in Android apps.

We have reviewed state-of-the-art papers in security and existing benchmarks for Android vulnerabilities, and identified twelve security code smells pertinent to ICC vulnerabilities.
In this paper we present these vulnerabilities and their corresponding smells in the code, and discuss how they could be eliminated or mitigated during development.
We present a lightweight static analysis tool on top of Android Lint that analyzes the code under development, and provides just-in-time feedback within the IDE about the presence of such security smells in the code.
Moreover, with the help of this tool we study the prevalence of security code smells in more than 700 open-source apps, and discuss the extent to which identifying these smells can uncover actual ICC security vulnerabilities.
We address the following three research questions:

\begin{itemize}

\item \textbf{RQ$_{1}$}:
\emph{What are the known ICC security code smells?}
We have reviewed related work, especially that appearing in top-tier conferences and journals, and identified twelve avoidable ICC vulnerabilities and the code smells that indicate their presence.
We discuss each smell, the risk associated with it, and its mitigation during app development.

\item \textbf{RQ$_{2}$}:
\emph{How prevalent are the smells in benign apps?}
We have developed a tool that statically analyzes apps for the existence of ICC security smells, and we applied it to a repository of 732 apps, mostly available on GitHub.
We discovered that almost all apps suffer from at least one category of ICC security smell, but fewer than 10\% suffer from more than two categories of such smells.
Interestingly, only small teams appear to be capable of consistently building software resistant to most security code smells.
Furthermore, long-lived projects have more issues than recently created ones, and updates rarely have any impact on ICC security.

\item \textbf{RQ$_{3}$}: 
\emph{To which extent does identifying security smells facilitate detection of security vulnerabilities?}
We inspected the identified smells in 100 apps, and verified whether they correspond to any security vulnerabilities.
Our investigation showed that about half of the identified smells are in fact good indicators of security vulnerabilities.

\end{itemize}

To summarize, this work represents an effort to spread awareness about the impact of programming choices in making apps secure, and to fundamentally reduce the attack surface of ICC APIs in Android.
We argue that this helps developers who develop security mechanisms to identify frequent problems, and also provides developers inexperienced in security with caveats about the prospect of security issues in their code.
Existing analysis tools often overwhelm developers with too many identified issues at once.
In contrast we provide feedback during app development where developers have the relevant context.
Such feedback makes it easier to react to issues, and helps developers to learn from their mistakes~\cite{Tymchuk:2018}.
This paper goes beyond our earlier work \cite{Ghafari:2017} by
\begin{inparaenum}[(i)]
\item providing a completely new study on ICC vulnerabilities, one of the most prevalent Android security issues, and identifying the corresponding security smells,
\item providing more precise, while still lightweight, static analysis tool support to identify such smells,
\item integrating our analysis into Android Lint, thus providing just-in-time feedback to developers, 
\item experimentation on a new dataset of open-source Android apps, and
\item open-sourcing the lint checks as well as the analyzed data.\footnote{We are collaborating with Google to officially integrate these checks into Android Studio.}
\end{inparaenum}

The remainder of this paper is organized as follows.
We provide the necessary background about the Android OS and ICC risks from which Android apps suffer in \autoref{sec:background}.
We introduce ICC-related security code smells in \autoref{sec:code-smells}, followed by our empirical study in \autoref{sec:empirical-study}.
We provide a brief overview of the related work in \autoref{sec:related-work}, before concluding the paper in \autoref{sec:conclusion}.

% ==================================================
\section{Background} \label{sec:background}

This section covers the necessary background in the Android platform, and briefly presents common security threats in the context of ICC scenarios.

% --------------------------------------------------
\subsection{Android Architecture}
Android is the most popular operating system (OS) for smartphones and other types of mobile devices.
It provides a rich set of APIs for app developers to access common features on mobile devices.

An Android app consists of an .apk file containing the compiled bytecode, any needed data, and resource files.
The Android platform assigns a unique user identifier (UID) to each app at installation time, and runs it in a unique process within a sandbox so that every app runs in isolation from other apps.
Moreover, access to sensitive APIs is protected by a set of permissions that the user can grant to an app.
In general, these permissions are text strings that correlate to a specific access grant, \eg \code{android.permission.CAMERA} for camera access.

Four types of components can exist in an app: activities, services, broadcast receivers, and content providers.
In a nutshell:

\begin{itemize}
\item \emph{Activities} build the user interface of an app, and allow users to interact with the app.

\item \emph{Services} run operations in the background, without a user interface.

\item \emph{Broadcast receivers} receive system-wide ``intents'', \ie descriptions of operations to be performed, sent to multiple apps.
Broadcast receivers act in the background, and often relay messages to activities or services.

\item \emph{Content providers} manage access to a repository of persistent data that could be used internally or shared between apps.
\end{itemize}

The OS and its apps, as well as components within the same or across multiple apps, communicate with each other via ICC APIs.
These APIs take an \emph{intent object} as a parameter.
An intent is either \emph{explicit} or \emph{implicit}.
In an explicit intent, the source component declares to which target component (\ie Class or ComponentName instances) the intent is sent, whereas in an implicit intent, the source component only specifies a general action to be performed (\ie represented by a text string), and the target component that will receive the intent is determined at runtime.
Intents can optionally carry additional data also called \emph{bundles}.
Components declare their ability to receive implicit intents using ``intent filters'', which allow developers to specify the kinds of actions a component supports.
If an intent matches any intent filter, it can be delivered to that component.

% --------------------------------------------------
\subsection{ICC Threats}
ICC not only significantly contributes to the development of collaborative apps, but it also poses a common attack surface. 
The ICC-related attacks that threaten Android apps are:

\begin{itemize}
\item \smell{Denial of Service} 
Unchecked exceptions that are not caught will usually cause an app to crash.
The risk is that a malicious app may exploit such programming errors, and perform an inter-process denial-of-service attack to drive the victim app into an unavailable state.

\item \smell{Intent Spoofing} 
In this scenario a malicious app sends forged intents to mislead a receiver app that would otherwise not expect intents from that app.

\item \smell{Intent Hijacking} 
This threat is similar to a man-in-the-middle attack where a malicious app, registered to receive intents, intercepts implicit intents before they reach the intended recipient, and without the knowledge of the intent's sender and receiver.
\end{itemize}

\noindent Two major consequences of ICC attacks are as follows:

\begin{itemize}
\item \smell{Privilege Escalation}
The security model in Android does not by default prevent an app with fewer permissions (low privilege) from accessing components of another app with more permissions (high privilege).
Therefore, a caller can escalate its permissions via other apps, and indirectly perform unauthorized actions through the callee.

\item \smell{Data Leak} 
A data leak occurs when private data leaves an app and is disclosed to an unauthorized recipient.
\end{itemize}

% ==================SECTION==============================
\section{ICC Security Code Smells} \label{sec:code-smells}
In this section we present the guidelines we followed to derive the security code smells from previous research.
Finally, we explain each security smell in detail.

\subsection{Literature Review} \label{sec:methodology}
Although Android security is a fairly new field, it is very active, and researchers in this area have published a large number of articles in the past few years.
In order to answer the first research question ({RQ$_{1}$}), and to draw a comprehensive picture of recent ICC smells and their corresponding vulnerabilities, our study builds on two pillars, \ie a literature review and a benchmark inspection.\newline

We were essentially interested in any paper that matches our scope, \ie explaining an ICC-related issue, and any countermeasures that involve ICC communication in Android.

For our analysis we considered multiple online repositories, such as IEEE Xplore and the ACM Digital Library, as well as the Google Scholar search engine.
In each repository we formulated a search query comprising \emph{Android}, \emph{ICC}, \emph{IPC} and any other security-related keywords such as \emph{security}, \emph{privacy}, \emph{vulnerability}, \emph{attack}, \emph{exploit}, \emph{breach}, \emph{leak}, \emph{threat}, \emph{risk}, \emph{compromise}, \emph{malicious}, \emph{adversary}, \emph{defence}, or \emph{protect}.
In addition to increase our potential coverage on Android security, we also collected all related publications in recent editions of well-known software engineering venues like the \emph{International Conference on Software Engineering (ICSE)}.
This search led initially to 358 publications.

In order to retrieve only relevant information that lies within our scope, \ie \emph{Android application level ICC security}, 
we first read the title and abstract, and if the paper was relevant we continued reading other parts.

This process led to the inclusion of 47 papers in our study. We recursively checked both citations and cited papers until no new related papers were found. This added six new relevant papers in our list that in the end contained 53 relevant papers for an in-depth study, out a total of 430 papers. During the whole process, which was undertaken by two authors of this paper, we resolved any disagreement by discussions. The list of included papers in this study is available on the GitHub page of the project.\footnote{\url{https://github.com/pgadient/AndroidLintSecurityChecks}}

We further studied the well-known DroidBench\footnote{\url{https://github.com/secure-software-engineering/DroidBench}} and Ghera\footnote{\url{https://bitbucket.org/secure-it-i/android-app-vulnerability-benchmarks}} benchmarks for our evaluation, both built with a focus on ICC.
We relied on their technical implementation, or description where possible, to extract the desired information, \ie issues under test, symptoms, and vulnerabilities.
The inspection of these two benchmarks served two different purposes: on the one hand we wanted to ensure there are no smells that we might have missed to include in our list. On the other hand, we wanted to rely on some ground truth while explaining and examining the vulnerability capabilities of the smells.

%%%%%
\subsection{List of Smells} \label{sec:list-of-smells}

We have identified twelve ICC security code smells that are listed in \autoref{tab:list-of-smells}.
For each smell we report the security \emph{issue} at stake, the potential security \emph{consequences} for users, the \emph{symptom} in the code (\ie the code smell), the \emph{detection} strategy that has been implemented by our tool for identifying the code smell, any \emph{limitations} of the detection strategy, and a recommended \emph{mitigation} strategy of the issue, principally for developers.

\begin{table}[!htbp]
	\centering
	\begin{tabular}{ c l | c l }
		\textbf{ID} & \textbf{Security code smells} & \textbf{ID} & \textbf{Security code smells}\\ \hline
		SM01 & Persisted Dynamic Permission & SM07 & Broken Service Permission\\ \hline
		SM02 & Custom Scheme Channel & SM08 & Insecure Path Permission\\ \hline
		SM03 & Incorrect Protection Level & SM09 & Broken Path Permission Precedence\\ \hline
		SM04 & Unauthorized Intent & SM10 & Unprotected Broadcast Receiver\\ \hline
		SM05 & Sticky Broadcast & SM11 & Implicit Pending Intent\\ \hline
		SM06 & Slack WebViewClient & SM12 & Common Task Affinity\\ \hline
	\end{tabular}
	\caption{The identified ICC security code smells}
	\label{tab:list-of-smells}
\end{table}

We mined this information from numerous publications and benchmark suites, but only few of these resources provided detailed information about a given security issue. Therefore we put in a great manual effort to provide a comprehensive description for each smell, while consulting other resources such as the official Android documentation and external experts.
For instance, authors who focused on vulnerability detection generally neglected the aspect of mitigation.
This is very problematic, since it is very common for ICC-related issues to share strong similarities with only subtle differences, \eg regular directed inter-app communication and broadcasts both rely on intents.
Furthermore, manifold vulnerability terms that are used in the literature insufficiently reflect the symptoms as they do not name the involved component, \eg ``Confused Deputy'' instead of ``Unauthorized Intent''.
Better naming conventions would greatly ease the understanding of security vulnerabilities.

\begin{itemize}
\item \smell{SM01: Persisted Dynamic Permission}
Android provides access to protected resources through a Uniform Resource Identifier (URI) to be granted at runtime.\\
\Issue Such dynamic access is intended to be temporary, but if the developer forgets to revoke a permission, the access grant becomes more durable than intended.\\
\Consequences the recipient of the granted access obtains long-term access to potentially sensitive data.\\
\Symptoms \texttt{Context.grantUriPermission()} is present in the code without a corresponding \code{Context.revokeUriPermission()} call.\\
\Detection We report the smell when we detect a permission being dynamically granted without any revocations in the app.\\
\Limitations Our implementation does not match a specific grant permission to its corresponding revocation.
We may therefore fail to detect a missing revocation if another revocation is present somewhere in the code.\\
\Mitigation Developers have to ensure that granted permissions are revoked when they are no longer needed.
They can also attach sensitive data to the intent instead of providing its URI.

\item \smell{SM02: Custom Scheme Channel}
A \emph{custom scheme} allows a developer to register an app for custom URIs, \eg URIs beginning with \code{myapp://}, throughout the operating system once the app is installed.
For example, the app could register an activity to respond to the URI via an intent filter in the manifest.
Therefore, users can access the associated activity by opening specific hyperlinks in a wide set of apps.\\
\Issue Any app is potentially able to register and handle any custom schemes used by other apps.\\
\Consequences malicious apps could access URIs containing access tokens or credentials, without any prospect for the caller to identify these leaks~\cite{Wang:2013b}.\\
\Symptoms If an app provides custom schemes, then a scheme handler exists in the manifest file or in the Android code.
If the app calls a custom scheme, there exists an intent containing a URI referring to a custom scheme.\\
\Detection The \code{android:scheme} attribute exists in the \code{intent-filter} node of the manifest file, or \code{IntentFilter.addDataScheme()} exists in the source code.\\
\Limitations We only check the symptoms related to receiving custom schemes.\\
\Mitigation Never send sensitive data, \eg access tokens via such URIs.
Instead of custom schemes, use system schemes that offer restrictions on the intended recipients.
The Android OS could maintain a verified list of apps and the schemes that are matched when there is such a call.

\item \smell{SM03: Incorrect Protection Level}
Android apps must request permission to access sensitive resources.
In addition, custom permissions may be introduced by developers to limit the scope of access to specific features that they provide based on the protection level given to other apps.
Depending on the feature, the system might grant the permission automatically without notifying the user, \ie signature level, or after the user approval during the app installation, \ie normal level, or may prompt the user to approve the permission at runtime, if the protection is at a dangerous level.\\
\Issue An app declaring a new permission may neglect the selection of the right protection level, \ie a level whose protection is appropriate with respect to the sensitivity of resources~\cite{Mitra:2017}.\\
\Consequences apps with inappropriate permissions can still use a protected feature.\\
\Symptoms Custom permissions are missing the right \code{android:protection\-Level} attribute in the manifest file.\\
\Detection We report missing protection level declarations for custom permissions.\\
\Limitations We cannot determine if the level specified for a protection level is in fact right.\\
\Mitigation Developers should protect sensitive features with dangerous or signature protection levels.

\item \smell{SM04: Unauthorized Intent}
Intents are popular as one way requests, \eg sending a mail, or requests with return values, \eg when requesting an image file from a photo library.
Intent receivers can demand custom permissions that clients have to obtain before they are allowed to communicate.
These intents and receivers are ``protected''.\\
\Issue Any app can send an unprotected intent without having the appropriate permission, or it can register itself to receive unprotected intents.\\
\Consequences apps could escalate their privileges by sending unprotected intents to privileged targets, \eg apps that provide elevated features such as camera access.
Also, malicious apps registered to receive implicit unprotected intents may relay intents while leaking or manipulating their data~\cite{Chin:2011}.\\
\Symptoms The existence of an unprotected implicit intent.
For intents requesting a return value, the lack of check for whether the sender has appropriate permissions to initiate an intent.\\
\Detection The existence of several methods on the \code{Context} class for initiating an unprotected implicit intent like \code{startActivity}, \code{sendBroadcast}, \code{send\-Ordered\-Broadcast}, \code{sendBroadcastAsUser}, and \code{send\-Ordered\-Broadcast\-As\-User}.\\
\Limitations We do not verify, for a given intent requesting a return value, if the sender enforces permission checks for the requested action.\\
\Mitigation Use explicit intents to send sensitive data.
When serving an intent, validate the input data from other components to ensure they are legitimate.
Adding custom permissions to implicit intents may raise the level of protection by involving the user in the process.

\item \smell{SM05: Sticky Broadcast}
A normal broadcast reaches the receivers it is intended for, then terminates.
However, a ``sticky'' broadcast stays around so that it can immediately notify other apps if they need the same information.\\
\Issue Any app can watch a broadcast, and particularly a sticky broadcast receiver can tamper with the broadcast~\cite{Mitra:2017}.\\
\Consequences a manipulated broadcast may mislead future recipients.\\
\Symptoms Broadcast calls that send a sticky broadcast appear in the code, and the related Android system permission exists in the manifest file.\\
\Detection We check for the existence of methods such as \code{send\-Sticky\-Broad\-cast}, \code{send\-StickyBroadcastAsUser}, \code{send\-Sticky\-Ordered\-Broad\-cast}, \code{send\-Sticky\-Ordered\-Broadcast\-As\-User}, \code{remove\-Sticky\-Broad\-cast}, and \code{remove\-Sticky\-Broad\-cast\-As\-User} on the \code{Context} object in the code and the \code{an\-droid.\-per\-mission.BROADCAST\_STICKY} permission in the manifest file.\\
\Limitations We are not aware of any limitations.\\
\Mitigation Prohibit sticky broadcasts.
Use a non-sticky broadcast to report that something has changed.
Use another mechanism, \eg an explicit intent, for apps to retrieve the current value whenever desired.

\item \smell{SM06: Slack WebViewClient}
A \code{WebView} is a component to facilitate web browsing within Android apps.
By default, a \code{WebView} will ask the Activity Manager to choose the proper handler for the URL.
If a \code{WebViewClient} is provided to the \code{WebView}, the host application handles the URL.\\
\Issue The default implementation of a \code{WebViewClient} does not restrict access to any web page~\cite{Mitra:2017}.\\
\Consequences it can be pointed to a malicious website that entails diverse attacks like phishing, cross-site scripting, \etc\\
\Symptoms The \code{WebView} responsible for URL handling does not perform adequate input validation.\\
\Detection The \code{WebView.setWeb\-View\-Client()} exists in the code but the \code{Web\-View\-Client} instance does not apply any access restrictions in \code{WebView.\-should\-Override\-Url\-Loading()}, \ie it returns \code{false} or calls \code{WebView.\-loadUrl()} right away.
Also, we report a smell if the implementation of \code{Web\-View.\-should\-Intercept\-Request()} returns \code{null}.\\
\Limitations It is inherently difficult to evaluate the quality of an existing input validation.\\
\Mitigation Use a white list of trusted websites for validation, and benefit from external services, \eg SafetyNet API,\footnote{\url{https://developer.android.com/training/safetynet/safebrowsing.html}} that provide information about the threat level of a website.

\item \smell{SM07: Broken Service Permission}
Two different mechanisms exist to start a service: \code{onBind} and \code{on\-Start\-Command}.
Only the latter allows services to run indefinitely in the background, even when the client disconnects.
An app that uses Android IPC to start a service may possess different permissions than the service provider itself.\\
\Issue When the callee is in possession of the required permissions, the caller will also get access to the service.\\
\Consequences a privilege escalation could occur~\cite{Mitra:2017}.\\
\Symptoms The lack of appropriate permission checks to ensure that the caller has access right to the service.\\
\Detection We report the smell when the caller uses \code{start\-Service}, and then the callee uses \code{check\-Calling\-Or\-Self\-Permis\-sion}, \code{en\-force\-Calling\-Or\-Self\-Per\-miss\-ion}, \code{check\-Call\-ing\-Or\-Self\-Uri\-Permis\-sion}, or \code{en\-force\-Cal\-ling\-Or\-Self\-Uri\-Per\-mis\-sion} to verify the permissions of the request.
Calls on the \code{Context} object for permission check will then fail as the system mistakenly considers the callee's permission instead of the caller's.
Furthermore, reported are calls to \code{check\-Permis\-sion}, \code{check\-Uri\-Permis\-sion}, \code{en\-force\-Permis\-sion}, or \code{en\-force\-Uri\-Permis\-sion} methods on the \code{Context} object, when additional calls to \code{get\-Cal\-ling\-Pid} or \code{get\-Cal\-ling\-Uid} on the \code{Binder} object exist.\\
\Limitations We currently do not distinguish between checks executed in \code{Ser\-vice.\-on\-Bind} or \code{Service.on\-Start\-Command}, and we do not verify custom permission checks based on the user id with \code{getCallingUid}.\\
\Mitigation Verify the caller's permissions every time before performing a privileged operation on its behalf using \code{Context.check\-Cal\-ling\-Permis\-sion()} or \code{Con\-text.\-check\-Call\-ing\-Uri\-Per\-mis\-sion()} checks.
If possible, do not implement \code{Service.on\-Start\-Command} in order to prevent clients from starting, instead of binding to, a service.
Ensure that appropriate permissions to access the service have been set in the manifest.

\item \smell{SM08: Insecure Path Permission}
Apps can access data provided by a content provider using path specifications of the form \code{/a/b/c}.
A content provider may restrict access to certain data under a given path by specifying so called path permissions.
For example, it may specify that other apps cannot access data located under \code{/data/secret}.
The Android framework prohibits access to unauthorized apps only if the requested path strictly matches the protected path.
For instance, \code{//data/secret} is different from \code{/data/secret}, and therefore the framework will not block access to it.\\ 
\Issue 
Developers often use the \code{UriMatcher} for URI comparison in the \code{query} method of a content provider to access data, but this matcher, unlike the Android framework, evaluates paths with two slashes as being equal to paths with one slash.\\
\Consequences access to presumably protected resources may be granted to unauthorized apps~\cite{Mitra:2017}.\\
\Symptoms A \code{UriMatcher.match()} is used for URI validation.\\
\Detection We look for \code{path-permission} attributes in the manifest file, and \code{UriMatcher.match()} methods in the code.\\
\Limitations We are not aware of any limitation.\\
\Mitigation As long as the bug exists in the Android framework, use your own URI matcher.

\item \smell{SM09: Broken Path Permission Precedence}
In a content provider, more fine-grained path permissions \eg on \code{/data/secret} take precedence over those with a larger scope \eg on \code{/data}.\\
\Issue A path permission never takes precedence over a permission on the whole content provider due to a bug that exists in the \code{ContentProvider.\-en\-force\-Read\-Permission\-Inner()} method.
For example, if a content provider has a permission for general use, as well as a path permission to protect \code{/data/secret} from untrusted apps, then the general use permission takes precedence.\\
\Consequences content providers may mistakenly grant untrusted apps access to presumably protected paths.\\
\Symptoms A content provider is protected by path-specific permissions.\\
\Detection We look for a \code{path-permission} in the definition of a content provider in the manifest file.\\
\Limitations We are not aware of any limitation.\\
\Mitigation
As long as the bug exists in Android, instead of path permissions use a distinct content provider with a dedicated permission for each path.

\item \smell{SM10: Unprotected Broadcast Receiver}
Static broadcast receivers are registered in the manifest file, and start even if an app is not currently running.
Dynamic broadcast receivers are registered at run time in Android code, and execute only if the app is running.\\
\Issue Any app can register itself to receive a broadcast, which exposes the app to any other app able to initiate the broadcast.\\
\Consequences if there is no permission check, the receiver may respond to a spoofed intent yielding unintended behavior or data leaks~\cite{Mitra:2017}.\\
\Symptoms The \code{Context.registerReceiver()} call without any argument for permission exists in the code\\
\Detection We report cases where the permission argument is missing or is \code{null}.\\
\Limitations We are not aware of the permissions' appropriateness.\\
\Mitigation Register broadcast receivers with sound permissions.

\item \smell{SM11: Implicit Pending Intent}
A \code{PendingIntent} is an intent that executes the specified action of an app in the future and on behalf of the app, \ie with the identity and permissions of the app that sends the intent, regardless of whether the app is running or not.\\
\Issue Any app can intercept an implicit pending intent~\cite{Mitra:2017} and use the pending intent's \code{send} method to submit arbitrary intents on behalf of the initial sender.\\
\Consequences a malicious app can tamper with the intent's data and perform custom actions with the permissions of the originator.
Relaying of pending intents could be used for intent spoofing attacks.\\
\Symptoms The initiation of an implicit \code{PendingIntent} in the code.\\
\Detection We report a smell if methods such as \code{getActivity}, \code{get\-Broadca\-st}, \code{getService}, and \code{getForegroundService} on the \code{PendingIntent} object are called, without specifying the target component\\
\Limitations Arrays of pending intents are not yet supported in our analysis.\\
\Mitigation Use explicit pending intents, as recommended by the official documentation.\footnote{\url{https://developer.android.com/reference/android/app/PendingIntent.html}}

\item \smell{SM12: Common Task Affinity}
A \emph{task} is a collection of activities that users interact with when carrying out a certain job.\footnote{\url{https://developer.android.com/guide/components/activities/tasks-and-back-stack.html}}
A task affinity, defined in the manifest file, can be set to an individual activity or at the application level.\\
\Issue Apps with identical task affinities can overlap each others' activities, \eg to fade in a voice record button on top of the phone call activity.
The default value does not protect the application against highjacking of UI components.\\
\Consequences malicious apps may hijack an app's activity paving the way for various kinds of spoofing attacks~\cite{Ren:2015}.\\
\Symptoms The task affinity is not empty.\\
\Detection We report a smell if the value of a task affinity is not empty.\\
\Limitations We are not aware of any limitation.\\
\Mitigation If a task affinity remains unused, it should always be set to an empty string on the application level.
Otherwise set the task affinity only for specific activities that are safe to share with others.
We suggest that Android set the default value for a task affinity to empty.
It may also add the possibility of setting a permission for a task affinity.

\end{itemize}

% --------------------------------------------------

In summary, each security smell introduces a different set of vulnerabilities.
We established a close relationship between the smells and the security risks with the purpose of providing accessible and actionable information to developers, as shown in \autoref{tab:relation-vul-smells}.

\begin{table}[!htbp]
	\centering
	\begin{tabular}{ l | l }
		\textbf{Vulnerabilities} & \textbf{Security code smells} \\ \hline
		Denial of Service & SM01, SM02, SM03, SM04, SM06, SM07, SM10, SM12 \\ \hline
		Intent Spoofing & SM02, SM03, SM04, SM05, SM07, SM08, SM09, SM10, SM11 \\ \hline
		Intent Hijacking & SM02, SM03, SM04, SM05, SM10, SM11 \\ \hline
	\end{tabular}
	\caption{The relationship between vulnerabilities and security code smells}
	\label{tab:relation-vul-smells}
\end{table}

% ==================================================
\section{Empirical Study} \label{sec:empirical-study}

In this section we first present the Lint-based tool with which we detect security code smells, and introduce a dataset of more than 700 open-source Android projects that are mostly hosted on GitHub.
We then present the results of our investigation into {RQ$_{2}$} and {RQ$_{3}$} by analyzing the prevalence of security smells in our dataset, and by discussing the performance of our tool, respectively.

The results in \autoref{sec:results} suggest that although fewer than 10\% of apps suffer from more than two categories of ICC security smells, only small teams are capable of consistently building software resistant to most security code smells.
With respect to app volatility, we discovered that updates rarely have any impact on ICC security, however, in case they do, they often correspond to new app features.
On the other hand, we found that long-lived projects have more issues than recently created ones, except for apps that receive frequent updates, where the opposite is true.
Moreover, the findings of Android Lint's security checks correlate to our detected security smells.

In \autoref{sec:manual-analysis}, our manual evaluation confirms that our tool successfully finds many different ICC security code smells, and that about 43.8\% of the smells in fact represent vulnerabilities.
We consequently hypothesize that the tool can offer valuable support in security audits, but this remains to be explored in our future work.

We performed analyses similar to our previous work, \eg exploring the relation between star rating and smells, or the distribution of smells in app categories, and we did not observe major differences with our past findings~\cite{Ghafari:2017}.
Our results are therefore in line with our prior research that did not consider ICC smells, and found that the majority of apps suffer from security smells, despite the diversity of apps in popularity, size, and release date.

% --------------------------------------------------
\subsection{Linting Tool}\label{sec:tool}
Our Linting tool is built using Android Lint, a static analysis framework from the official Android Studio IDE\footnote{\url{https://sites.google.com/a/android.com/tools/tips/lint}} for analyzing Android apps.
Android Lint provides various rich interfaces for analyzing XML, Java, and Class files in Android.
Using these interfaces, one can implement a so-called ``detector'' that is responsible for scanning code, detecting issues, and reporting them.
More specifically, each detector is represented by a Java class that implements Android Lint interfaces to access Android Lint's abstract syntax trees (ASTs) of the app built from XML, source, or bytecode.
In order to ease the AST traversal, Android Lint provides an implementation of the visitor design pattern with additional helper methods to support further interaction with the tree.
The majority of methods use idiomatic names that closely resemble the developer's intention, \eg \code{Uast\-Utils.\-try\-Resol\-ve()} to resolve a variable, or the class \code{Const\-ant\-Eval\-uator} to evaluate constants.
The latest Android Lint provides more than 300 different detectors to check several categories of issues such as, \eg Accessibility, Usability, Security, \etc

We extended Android Lint by developing twelve new detectors.
These detectors implement \code{UastScanner}\footnote{The \code{UastScanner} is the successor of the \code{JavaScanner}, and, in addition to Java, also supports Kotlin, a new programming language used in the Android platform.} and \code{XmlScanner} interfaces to check the presence of security code smells in source code and manifest files, respectively.
We implemented the detection strategies that we introduced for each security smell in \autoref{sec:code-smells}.
The complexity of our detectors varies; the average size of a detector is 115 lines of code.

Android Lint brings analysis support directly into the Android Studio IDE.
Developers can therefore receive just-in-time feedback during app development about the presence of security code smells in their code.
For this purpose, the .jar file that contains our detectors should be copied into the Lint directory.
These detectors will then be run automatically during programming in the latest Android Studio IDE (\ie the Canary build), and notify developers about the security code smells once they appear in the code under development.
Each notification includes an explanation of the smell, mitigation or elimination strategies, as well as a link to some references.

Linting in batch mode is also possible through the command line interface, given the availability of the successfully built projects.
In our experience, a successful build often entails changing build paths, and updating Gradle and its project configurations to a version that is compatible with the current release of Android Lint.
We created a script to automate most of this non-trivial process.
After a successful build of each project, another script runs the executable of Android Lint, and collects the analysis results in XML files.

The tool is publicly available for download from a GitHub repository.\footnote{\url{https://github.com/pgadient/AndroidLintSecurityChecks}}

% --------------------------------------------------
\subsection{Dataset}\label{sec:dataset}
% Dataset
We collected all open-source apps from the F-Droid\footnote{\url{https://f-droid.org/}} repository as well as several other apps directly from GitHub.\footnote{\url{https://github.com/pcqpcq/open-source-android-apps}}
In total we collected 3\,471 apps, of which we could successfully build 1\,487 (42\%).
For replication of our results we explicitly provide the package names of all successfully analyzed apps,\footnote{\url{https://github.com/pgadient/AndroidLintSecurityChecks/blob/master/dataset/analyzed_apps.csv}} instead of a binary compilation, because of the dataset's storage space requirements of more than 27 GBytes.
In order to reduce the influence of individual projects, in case there existed more than one release of a project, we only considered the latest one.
Finally, we were left with 732 apps (21\%) in our dataset.
The median project size in our dataset is about 1.2 MB, while the median number of data files per project is 108.

% --------------------------------------------------
\subsection{Batch Analysis}
\label{sec:results}

This section presents the results of applying our tool to all the apps in our dataset.

% --------------------------------------------------
\subsubsection{Prevalence of Security Smells}
\begin{figure}
\includegraphics[width=\columnwidth, trim = 2cm 2cm 2cm 2cm]{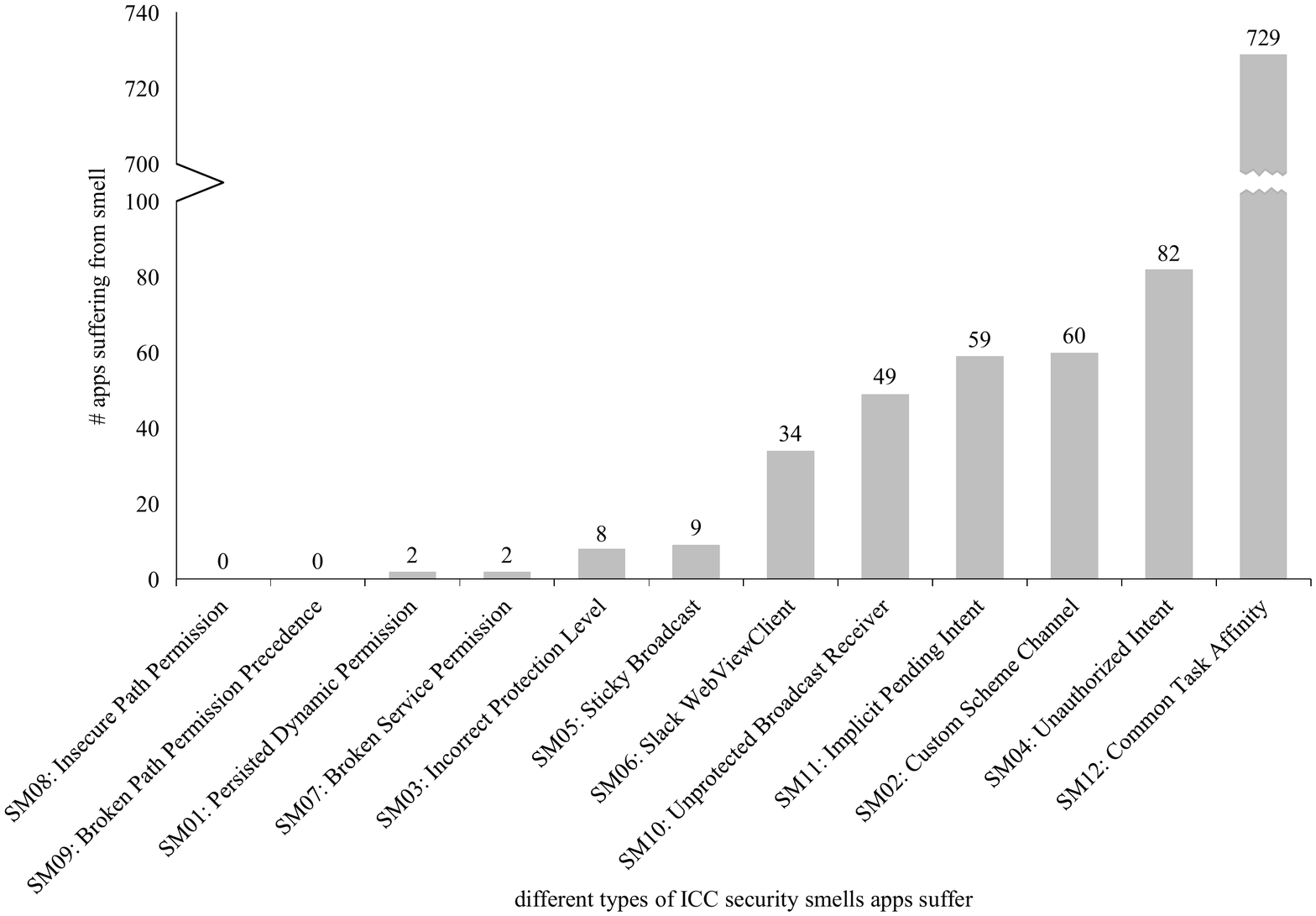}
\caption{Distribution of security smells in the apps}
\label{p_LR_smellprevalence_app}
\end{figure}

\autoref{p_LR_smellprevalence_app} shows how prevalent the smells are in our dataset.
Almost all apps suffer from \emph{Common Task Affinity} issues (99\%) followed by the much less prevalent \emph{Unauthorized Intent} smell (11\%).
The default value of task affinity configurations does not protect the application against highjacking of UI components, and only few developers appear to be aware of the issue and set the property accordingly.
\emph{Custom Scheme Channel} and \emph{Implicit Pending Intent} each contribute about 8\% of the smells.
Furthermore, \emph{WebViewClient} is in line with our observation that apps increasingly rely on web components for their UI.
At the other end of the spectrum, \emph{Sticky Broadcast}, \emph{Incorrect Protection Level}, \emph{Broken Service Permission}, and \emph{Persisted Dynamic Permission} cause less than 2\% of all issues.
The threat of path permissions is not very common, as no apps suffered from SM08 or SM09.

\begin{figure}
\includegraphics[width=\columnwidth, trim = 2cm 2cm 2cm 2cm]{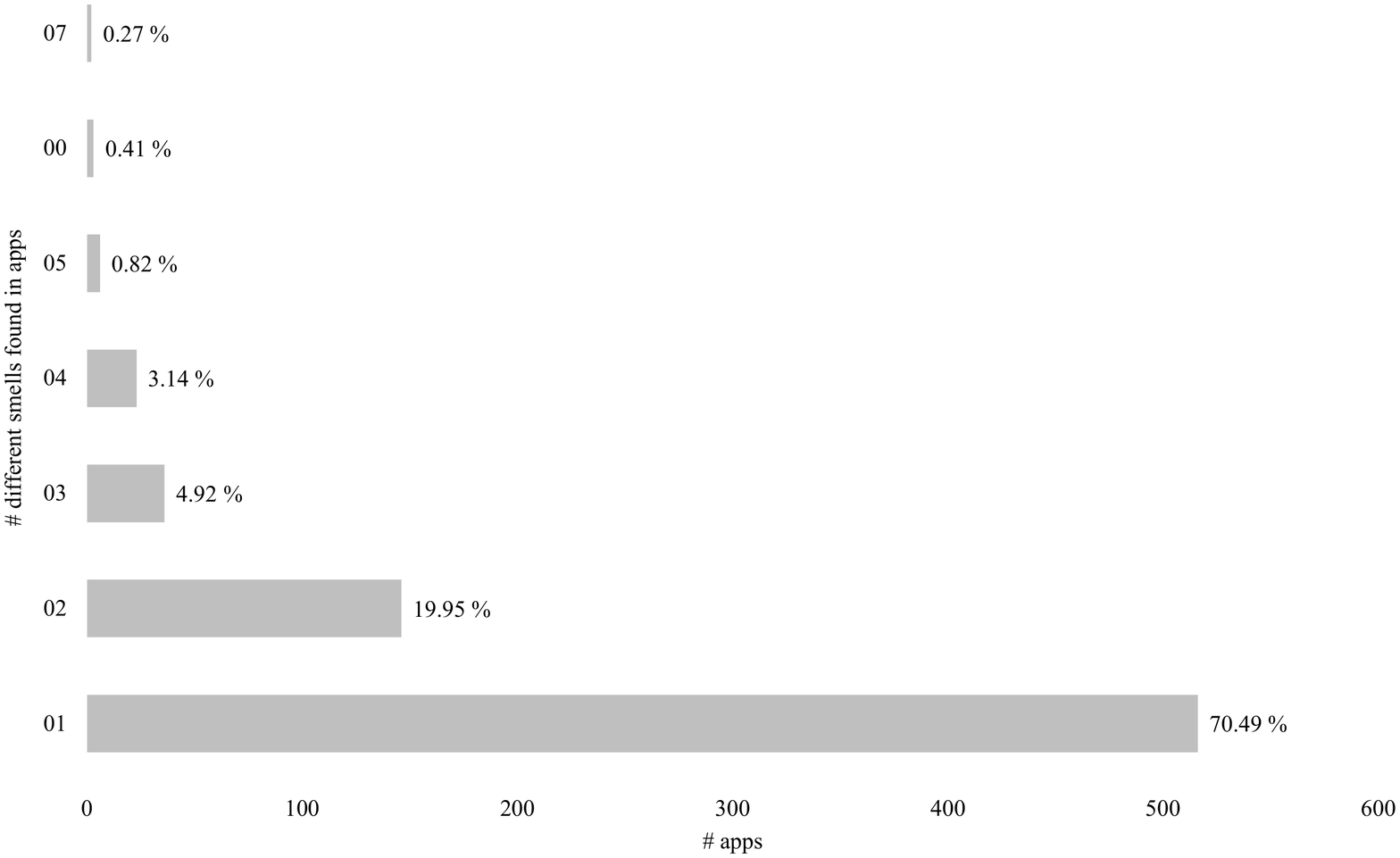}
\caption{Prevalence of different security smells in apps}
\label{p_LR_smelldistribution}
\end{figure}

We were also interested in the relative prevalence of different security smells in the apps (see \autoref{p_LR_smelldistribution}).
Less than 1\% did not suffer from any security smell at all, whereas the majority of apps, \ie over 90\%, suffered from one or two different smells.
9\% of all apps were affected by three or more smells.
No apps, fortunately, suffered from more than seven different types of smells.
It is important to recall that the more issues that are present in a benign app, the more likely it is that a malign app can exploit it, \eg with denial of service, intent spoofing, or intent hijacking attacks.

% --------------------------------------------------

\subsubsection{Contributor Affiliation}
\begin{figure}
\includegraphics[width=\columnwidth, trim = 2cm 2cm 2cm 2cm]{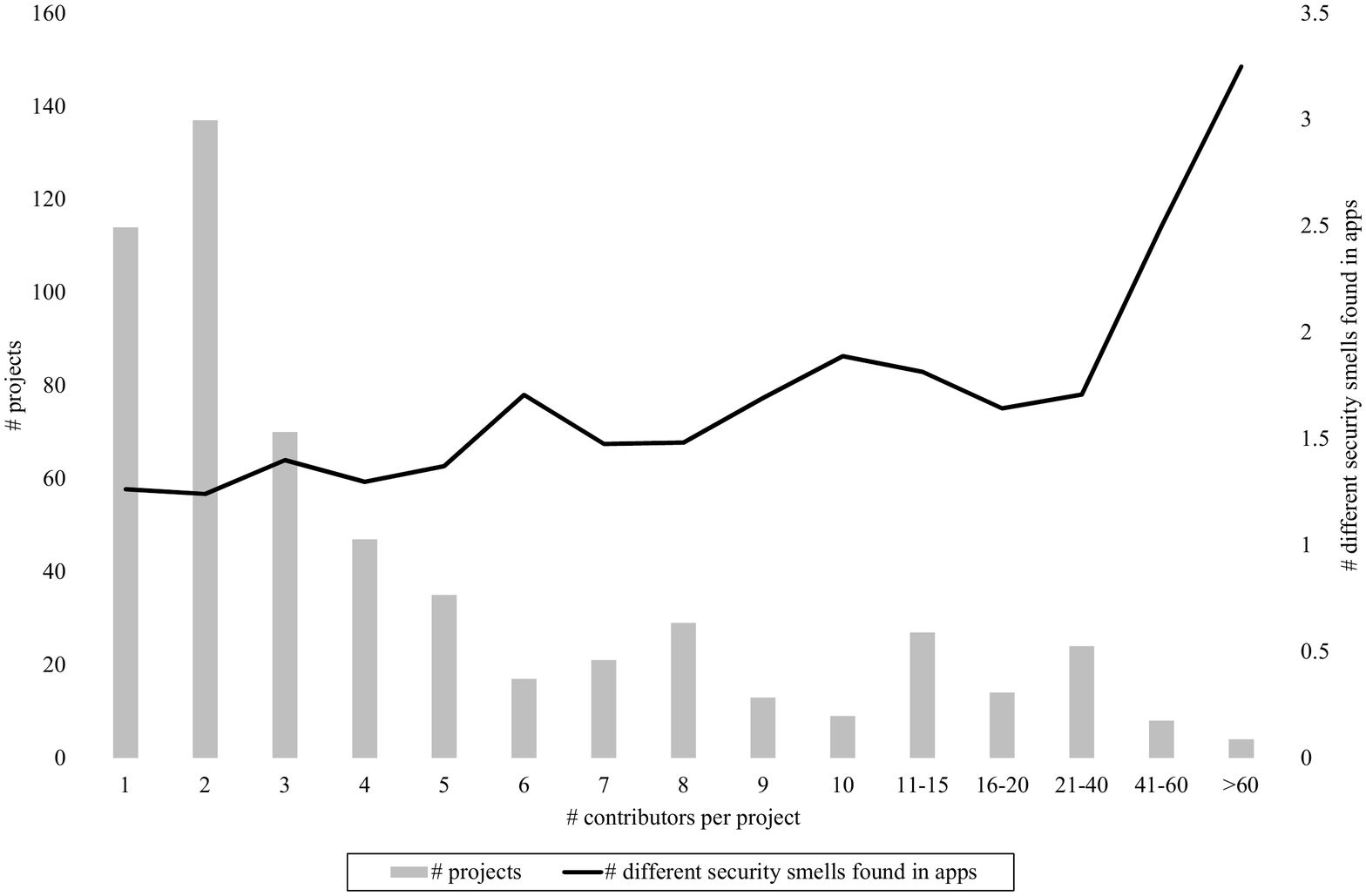}
\caption{Relation between number of a project's participants, its prevalence, and the average number of different security smells found}
\label{p_devs_per_project}
\end{figure}

\autoref{p_devs_per_project} shows the relationship between the number of contributors participating in a project and the mean number of security smell categories apps suffer from.
For example, the second last bar represents the number of all projects maintained by 41 to 60 participants, while the line chart shows that projects with this many participants suffer on average from 2.5 security smell categories.
We see that most apps are maintained by two contributors, followed by projects developed by individuals.
A trend exists that projects with many participants are less common than projects with only a few contributors.
The more people are involved in a project the more the security decreases, especially for large teams.
More precisely, we found statistical evidence that only small teams of up to five people are capable of consistently building projects resistant to most security code smells, by using the nonparametric Mann-Whitney U test that does not require the assumption of normal distributions for the dataset.
The mean different smell occurrences in the groups ``projects with one contributor'' and ``projects with six contributors'' were 1.263 and 1.705; the distributions in the two groups differed significantly (Mann-Whitney U = -2.086, n1 < n2 = 0, P < 0.05 two-tailed).
Similarly, we found that the distributions in the two groups ``projects with six to forty contributors'' and ``projects with more than forty contributors'' were diverse (Mann-Whitney U = -2.204, n1 < n2 = 0, P < 0.05 two-tailed) with mean different smell occurrences of 1.655 and 2.750, respectively.

% --------------------------------------------------

\subsubsection{App Updates}

We investigated the smell occurrences in subsequent app releases.
Of the 732 projects, 33 (4\%) of them released updates that either resolved or introduced issues.
By inspection of source code we noticed that many of the updates targeted new functionality, \eg addition of new implicit intents to share data with other apps, implementation of new notification mechanisms for receiving events from other apps using implicit pending intents, or registration of new custom schemes to provide further integration of app related web content into the Android system.
We believe this is due to developers focusing on new features instead of security.

For the majority of the app updates that introduced new security smells, we found that the dominant cause for decreased security is the accommodation of social interactions and data sharing features in the apps updates.
Hence, developers should be particularly cautious when integrating new functionality into an app.

% --------------------------------------------------
\subsubsection{Evolution}

Every new Android version introduces changes that strengthen security. The targeting of outdated Android releases will not only limit the supported feature set to the respective  release, but also introduce potential security issues as security fixes are continuously integrated into the OS with each update.

\begin{figure}
\includegraphics[width=\columnwidth, trim = 2cm 2cm 2cm 2cm]{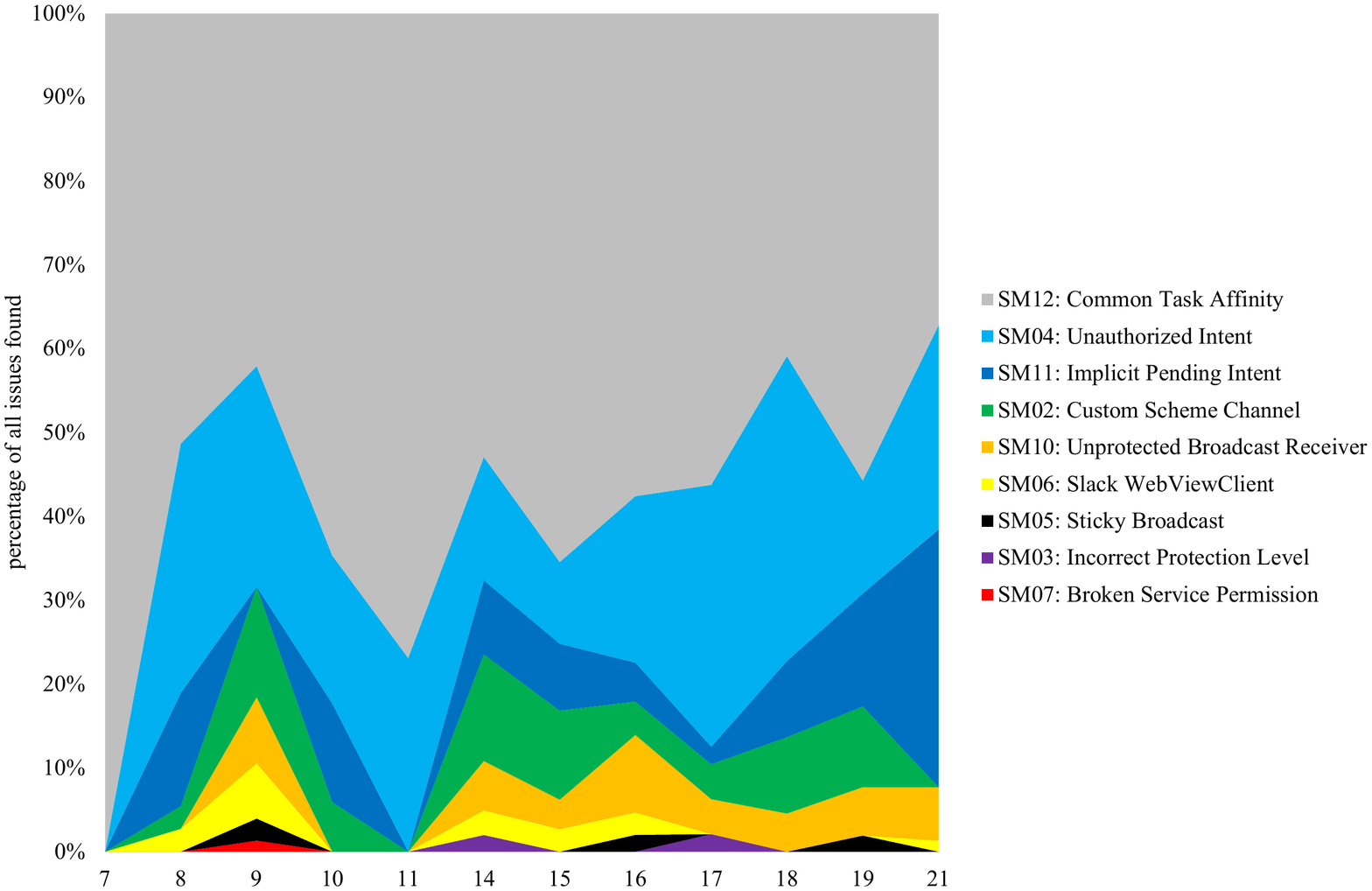}
\caption{Evolution of security code smells in different Android releases}
\label{vuln_evol_dist_p}
\end{figure}

\autoref{vuln_evol_dist_p} shows the evolution of security smells across different Android releases.
For those apps that had more than one release in our dataset, we only considered the latest release.
The horizontal axis shows the different Android releases apps are targeting in their configuration, whereas the vertical axis shows the contribution of a specific smell to the total amount of smells detected.
As in previous work~\cite{Ghafari:2017}, we see changes in some of the security smells apps suffer from.
We believe that the positive trend in \emph{Unauthorized Intent} within apps is the consequence of built-in sharing functionalities to external services.
The relative growth of \emph{Implicit Pending Intent} could correlate to the introduction of a new storage access framework in Android release 19, which heavily relies on intents, and allows developers to browse and open documents, images, and other files with ease.
Google's efforts to raise the developer's awareness of web-related security issues appears to be working: the occurrences of \emph{Slack WebView Client} have decreased in more recent releases.
Despite the lack of comprehensive data on API levels 10 and 11 due to the relatively few apps available for study, the occurrences of the majority of smells remain constant as a result of the early feature availability since API level 1.

% --------------------------------------------------
\subsubsection{Comparison to Existing Android Lint Checks}

In order to compare our findings with other issues in the apps, we correlated the results from the existing Android Lint framework with security code smells.
We wanted to explore whether frequent reports of specific Android Lint issue categories were also indicative of security issues, or in other words, if security checks by the Android Lint framework agree with our security smells and whether other quality aspects of an app could relate to its security level.
We collected all available issue reports for each app and then extracted the occurrences of each detected issue.

We applied the Pearson product-moment correlation coefficient algorithm for each ICC security smell category combination according to the following formula:
\newcommand*\mean[1]{\bar{#1}}

\small
\begin{equation*}
\begin{split}
&Pearson(x,y)=\frac{\sum(x-\mean{x})(y-\mean{y})}{\sqrt{\sum(x-\mean{x})^2 \sum(y-\mean{y})^2}}, \text{where}\\
\end{split}
\end{equation*}
\begin{equation*}
\begin{split}
&x \text{ is the array of all apps issue occurrences in category \emph{ICC security code smells}},\\
&y \text{ is the array of all apps issue occurrences in the respective Android Lint category}, \text{and}\\
&\mean{x},\mean{y} \text{ represent the corresponding sample means}. 
\end{split}
\end{equation*}
\normalsize

It provides a linear correlation between two vectors represented as a value in the range of $-1$ (total negative linear correlation) and $+1$ (total positive linear correlation).
The correlation of the Android Lint categories and our ICC smell category in \autoref{tab:correlation_matrix} reveals several interesting findings:
\begin{inparaenum}[(1)]
\item Our ICC security category strongly correlates with the Android Lint security category (+0.72), which contains checks for a variety of security-related issues such as the use of user names and passwords in strings, improper cryptography parameters, and bypassed certificate checks in WebView components. 
\item Another discovery is the minor correlation between the ICC security smells and the Android Lint correctness category (+0.29).
This category includes checks for erroneously configured project build parameters, incomplete view layout definitions, and usages of deprecated resources. 
\item Furthermore, we assume that usability does not impede security (+0.07), because issues in usability are closely related to UI mechanics.

\item Finally, minor correlations are shown for performance, accessibility, and internationalization.
These three categories have in common that they rely heavily on UI controls and configurations.
\end{inparaenum}

\definecolor{colorA}{RGB}{249,212,216} % Correctness
\definecolor{colorB}{RGB}{249,202,204} % Correctness: Messages
\definecolor{colorC}{RGB}{107,193,130} % Security
\definecolor{colorD}{RGB}{249,193,196} % Performance
\definecolor{colorE}{RGB}{249,172,174} % Usability: Typography
\definecolor{colorF}{RGB}{247,121,124} % Usability: Icons
\definecolor{colorG}{RGB}{247,105,107} % Usability
\definecolor{colorH}{RGB}{249,191,195} % Accessibility
\definecolor{colorI}{RGB}{247,133,135} % Internationalization
\definecolor{colorJ}{RGB}{247,121,124} % Internationalization: Bidirectional
\begin{table}[!htbp]
\centering
\begin{tabular}{ l | c }
\textbf{Android Lint category} & \textbf{Correlation with ICC security smells} \\ \hline
Security & \cellcolor{colorC}0.72 \\ \hline
Correctness & \cellcolor{colorA}0.29 \\ \hline
Correctness: Messages & \cellcolor{colorB}0.27 \\ \hline
Accessibility & \cellcolor{colorH}0.25 \\ \hline
Performance & \cellcolor{colorD}0.25 \\ \hline
Usability: Typography & \cellcolor{colorE}0.21 \\ \hline
Internationalization & \cellcolor{colorI}0.13 \\ \hline
Internationalization: Bidirectional & \cellcolor{colorJ}0.11 \\ \hline
Usability: Icons & \cellcolor{colorF}0.11 \\ \hline
Usability & \cellcolor{colorG}0.07 \\ \hline
\end{tabular}
\caption{Correlation of ICC security smells with Android Lint issue categories}
\label{tab:correlation_matrix}
\end{table}

\begin{figure}
\centering
\begin{tabular}{@{}c@{}}
\includegraphics[width=\columnwidth, trim = 2cm 2cm 2cm 2cm]{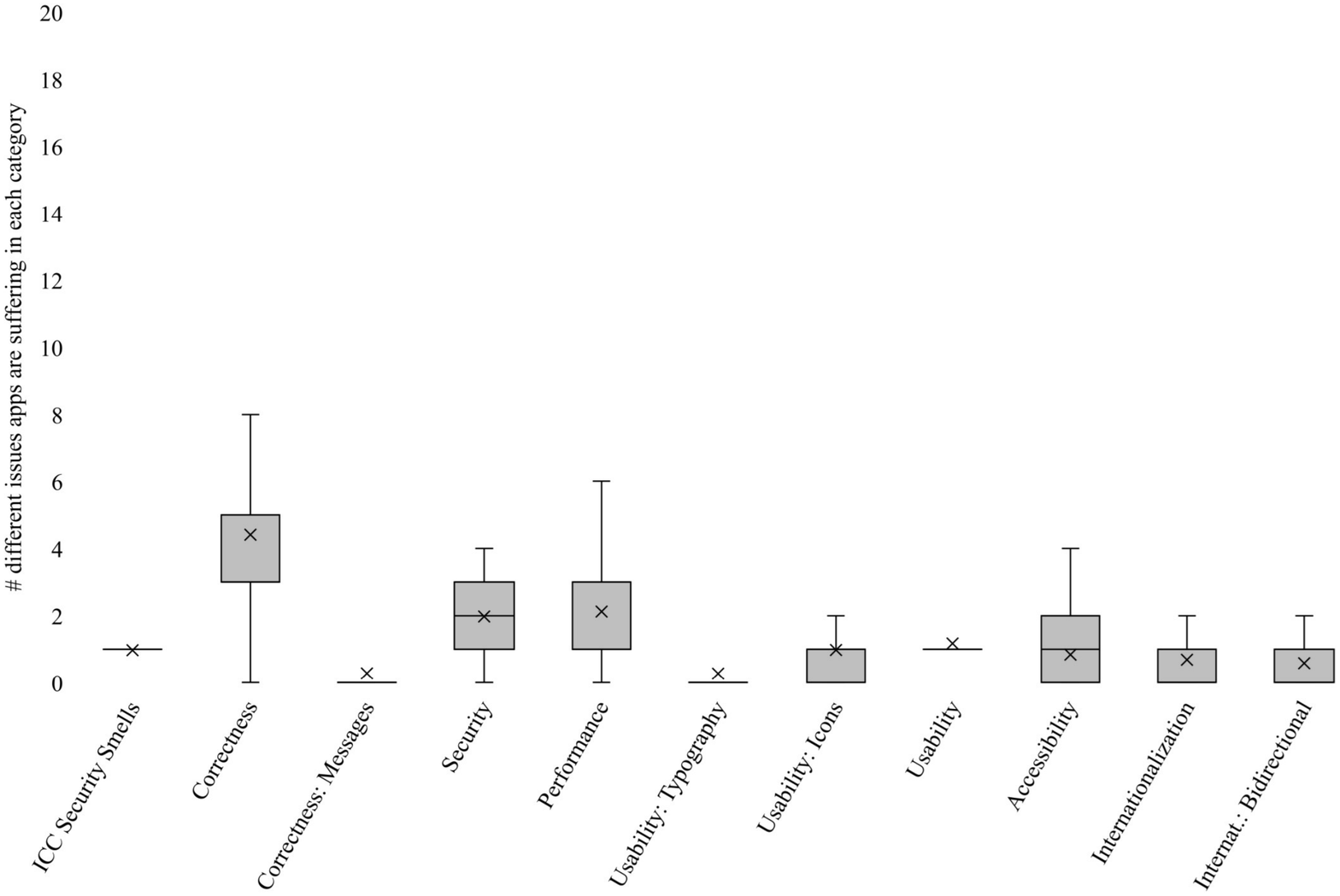}\\
\small (a) 100 least vulnerable apps
\end{tabular}

\vspace{\floatsep}

\begin{tabular}{@{}c@{}}
\includegraphics[width=\columnwidth, trim = 2cm 2cm 2cm 2cm]{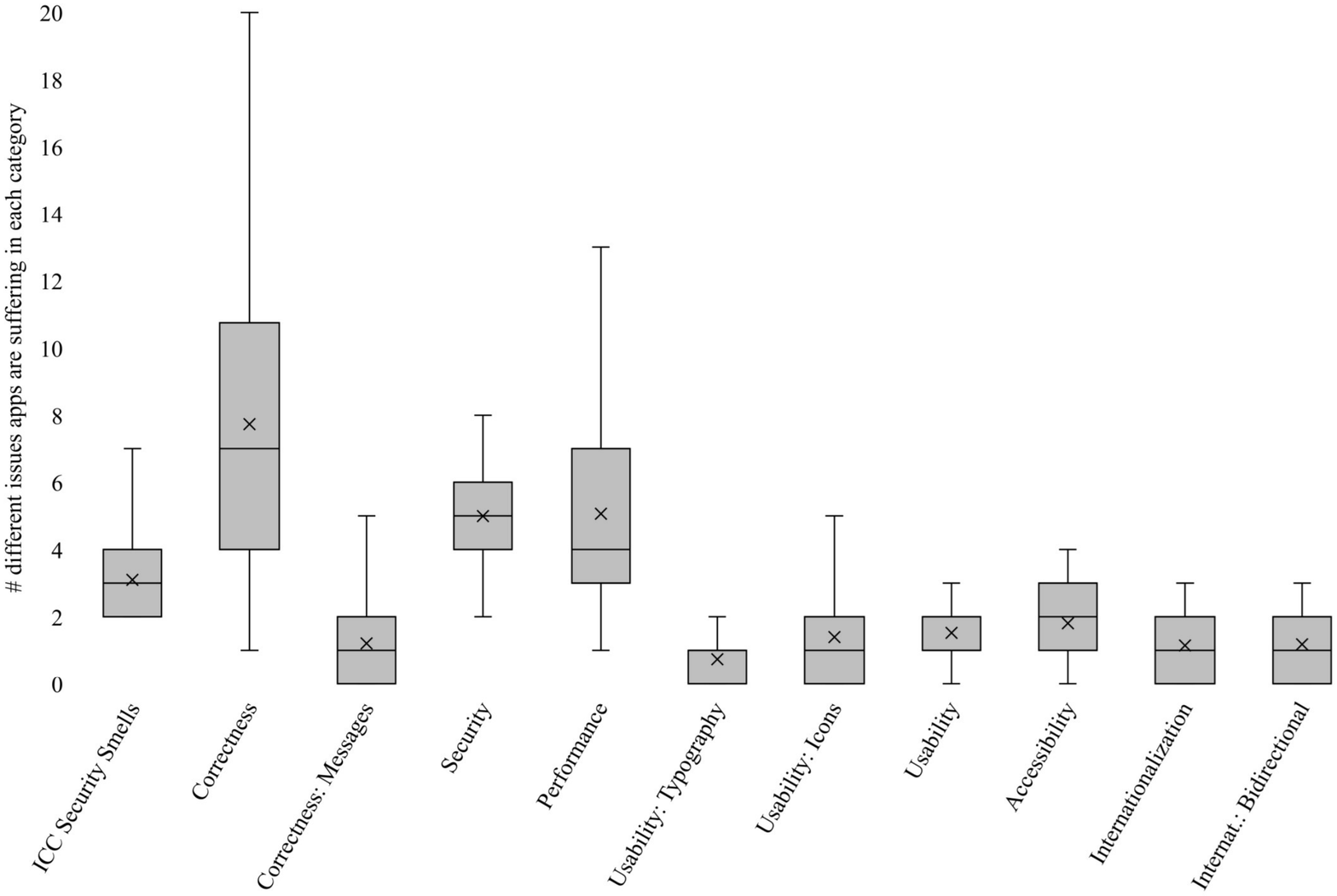}\\
\small (b) 100 most vulnerable apps
\end{tabular}

\caption{Prevalence of Android Lint issues in the 100 most and least vulnerable apps}
\label{tab:lint_distr_combined}
\end{figure}

To further assess how our tool performs on real world apps against the Android Lint detections, we take the 100 apps with the most and least prevalent ICC security smells and compare them to Android Lint's analysis results.
We expect to see significant similarities in the increase of issues detected as our security smells correlate to Android Lint's security checks, \ie the least vulnerable apps should suffer less in both, the Android Lint checks and our security smell detectors.
\autoref{tab:lint_distr_combined} illustrates two plots, each presenting our analysis results for the 100 apps suffering the most and the least from ICC security smells, respectively.
The vertical axis represents the condensed mean number of found issues, that is, we conflated all detected ICC security smell issues, regardless of their smell categories, into ``ICC Security Smells''.
The remaining Android Lint categories on the x-axis are treated accordingly.
The crosses represent the mean value of the number of different issues apps are suffering from in each category, and, as we hid any outliers to increase readability, these values can exceed the first quartiles.
The least and most affected apps clearly correspond in terms of issue frequency among specific categories, that is, the mean number of issues found in \emph{each} category is between 29\% and 332\% higher on behalf of the 100 most vulnerable apps.
Besides the ICC Security Smells category with an increase of 219\% in issues found, the Android Lint security category experienced an increase of 152\%.
The Correctness: Message and the Usability: Typography categories of Android Lint achieved, unexpectedly, an increase in issues found of about 332\% and 174\%, respectively.
After manual verification, we discovered that these gains were mostly caused by flawed language dictionary entries used for internationalization, such as missing or misunderstood language dependent string declarations, spelling mistakes, and the use of strings containing three dots instead of the ellipsis character.
While the 100 most vulnerable apps appear to prominently incorporate translations for several different languages, the 100 least vulnerable apps rarely make use of these features, hence, they suffer from much fewer issues.
The remaining categories encountered an increase of less than 139\%.
Interestingly, the internationalization category does not encounter a noticeable increase in issues due to its limited scope, \ie it only covers five specific flaws regarding insufficient language adaption, and the use of uncommon characters or encodings.
We propose that some of these issue detections should be reallocated to other categories, \eg spelling mistakes should be assigned to internationalization, and vice versa the issue \emph{SetTextI18n} in the category inter\-nation\-al\-ization that reports any use of methods that potentially fail with number conversions.

% --------------------------------------------------
\subsubsection{Influence of Project Age and Activity}
To explore the effect of recent updates, which we believed would improve app security, we evaluated our ICC category as well as the Android Lint security and correctness categories according to time since the last commit.
More precisely, we were interested in the question: Do recent updates improve app security?
A related question arises from the age of a project, \ie are mature projects more secure than recent creations?
We investigated these two questions based on available GitHub metadata, and brought the dates into perspective with the reported issues.

\begin{figure}
\centering
\begin{tabular}{@{}c@{}}
% config for three plots on a page: width=8.25cm, trim = 2cm 2cm 2cm 2cm
\includegraphics[width=\columnwidth, trim = 2cm 2cm 2cm 2cm]{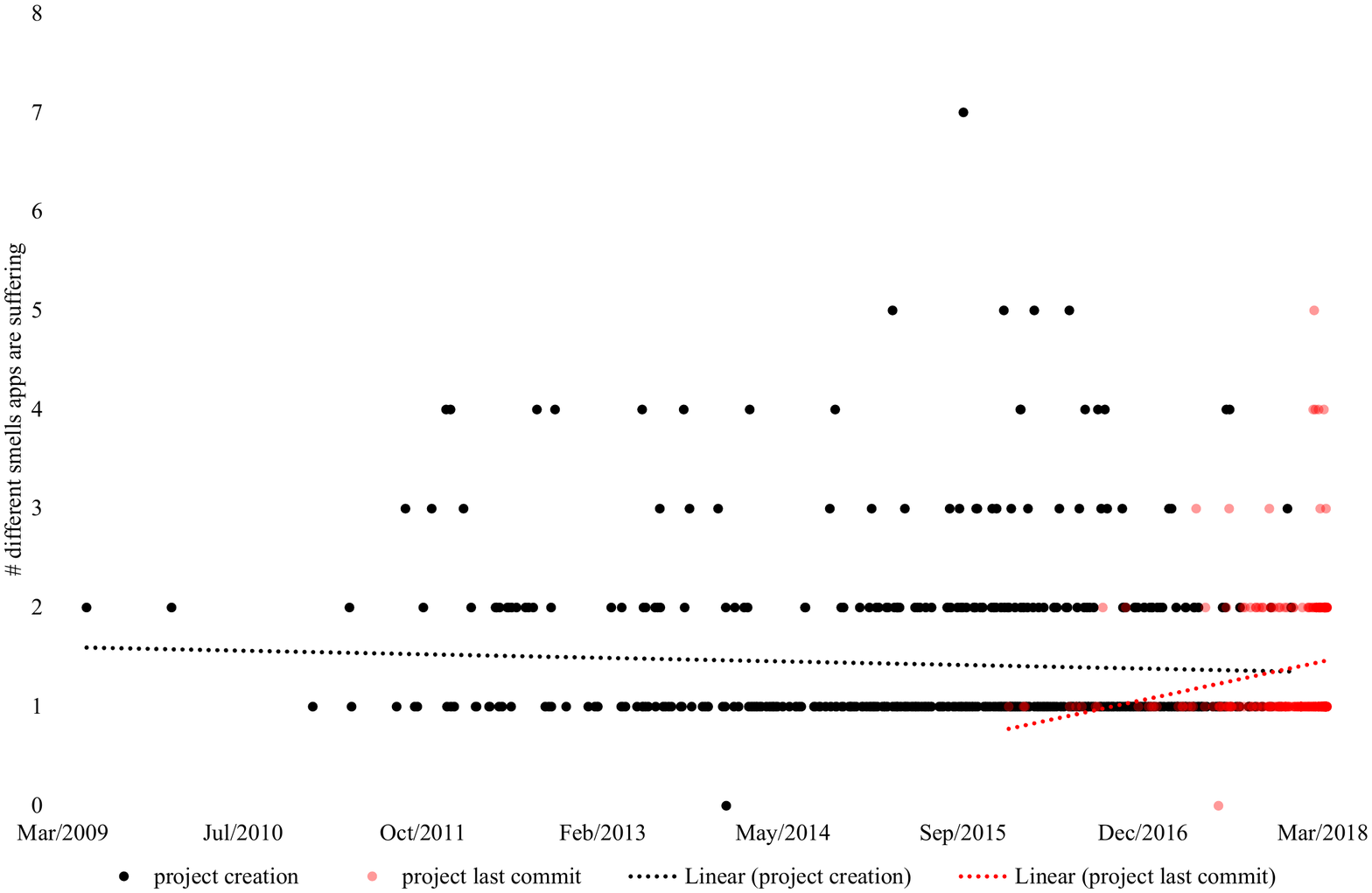}\\
\small (a) Relation of dates to our ICC security smells
\end{tabular}

\vspace{\floatsep}
\vspace{\floatsep}
\vspace{\floatsep}
\vspace{\floatsep}
\vspace{\floatsep}

\begin{tabular}{@{}c@{}}
% config for three plots on a page: width=8.25cm, trim = 2cm 2cm 2cm 2cm
\includegraphics[width=\columnwidth, trim = 2cm 2cm 2cm 2cm]{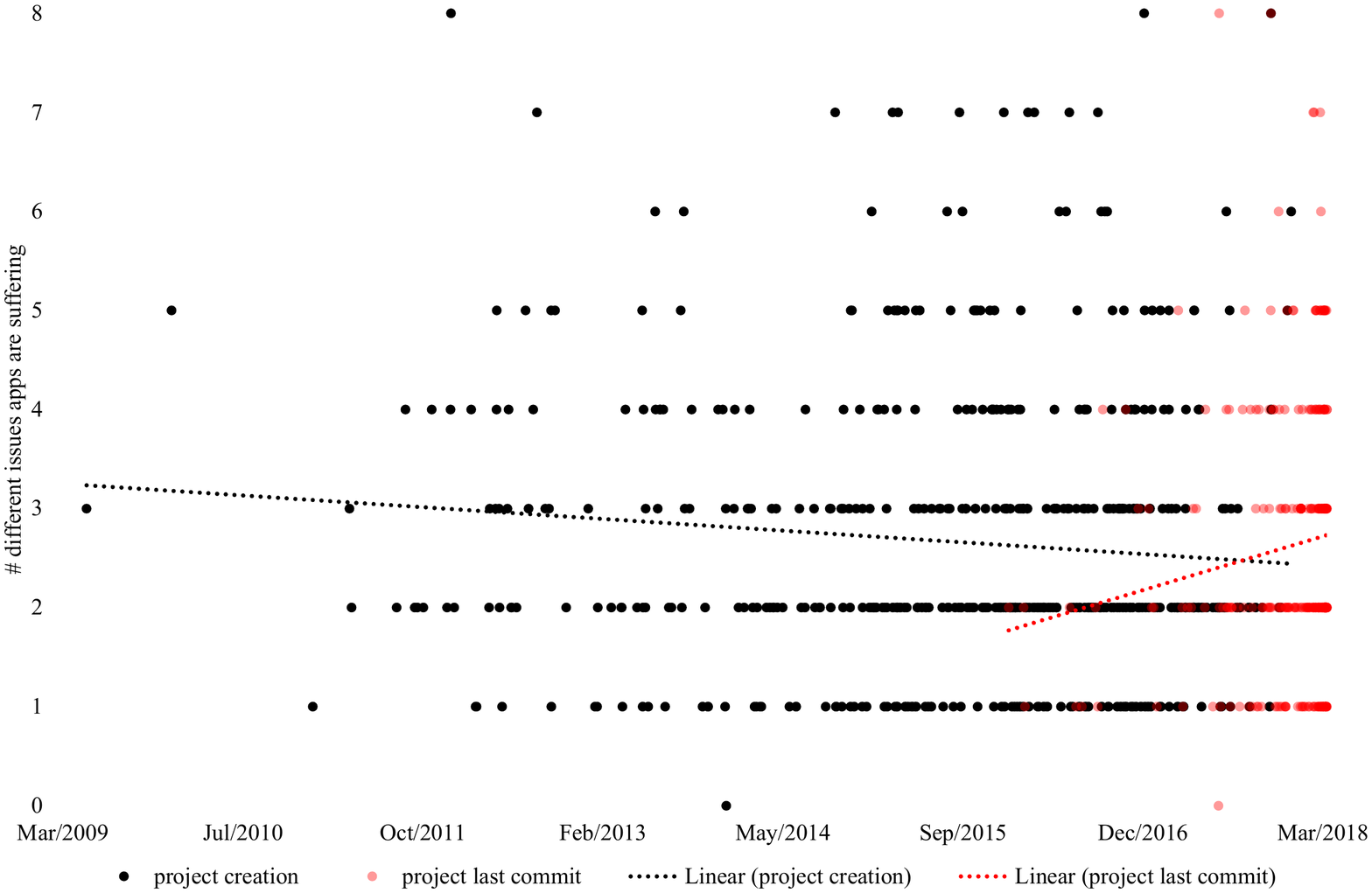}\\
\small (b) Relation of dates to Lint security
\end{tabular}

\caption{GitHub project creation and last commit date in relation to each project's issue count}
\label{lint_creation_update_data}
\end{figure}

\autoref{lint_creation_update_data} shows the mean number of detected issues per app on the vertical axis, either for the ICC security smells, or the Android Lint security category.
The black dots reveal the app's project creation dates, whereas red dots indicate the most recent commit dates of projects, hence every app is represented by one black and one red dot in each plot.
The creation date for the majority of apps dates back to less than 6.5 years.
We can clearly see in every plot a correlation between both the creation date, and the date of the last commit to the overall issue count, based on the pictured linear trends (dotted lines). 
These trends, which are very similar in terms of elevation, are a further indicator for the close relationship between our tool and the Android Lint checks.
Moreover, the Lint security category shows strong evidence that mature projects have more security issues than recent ones.
We assume that this is caused by the less comprehensive checks that older IDEs performed on the source code.
Similarly, apps that frequently introduce changes, \ie receive updates, are prone to have more issues.

\subsubsection{Influence of Code Size}
\begin{figure}
\centering
\begin{tabular}{@{}c@{}}
% config for three plots on a page: width=8.25cm, trim = 2cm 2cm 2cm 2cm
\includegraphics[width=\columnwidth, trim = 2cm 2cm 2cm 2cm]{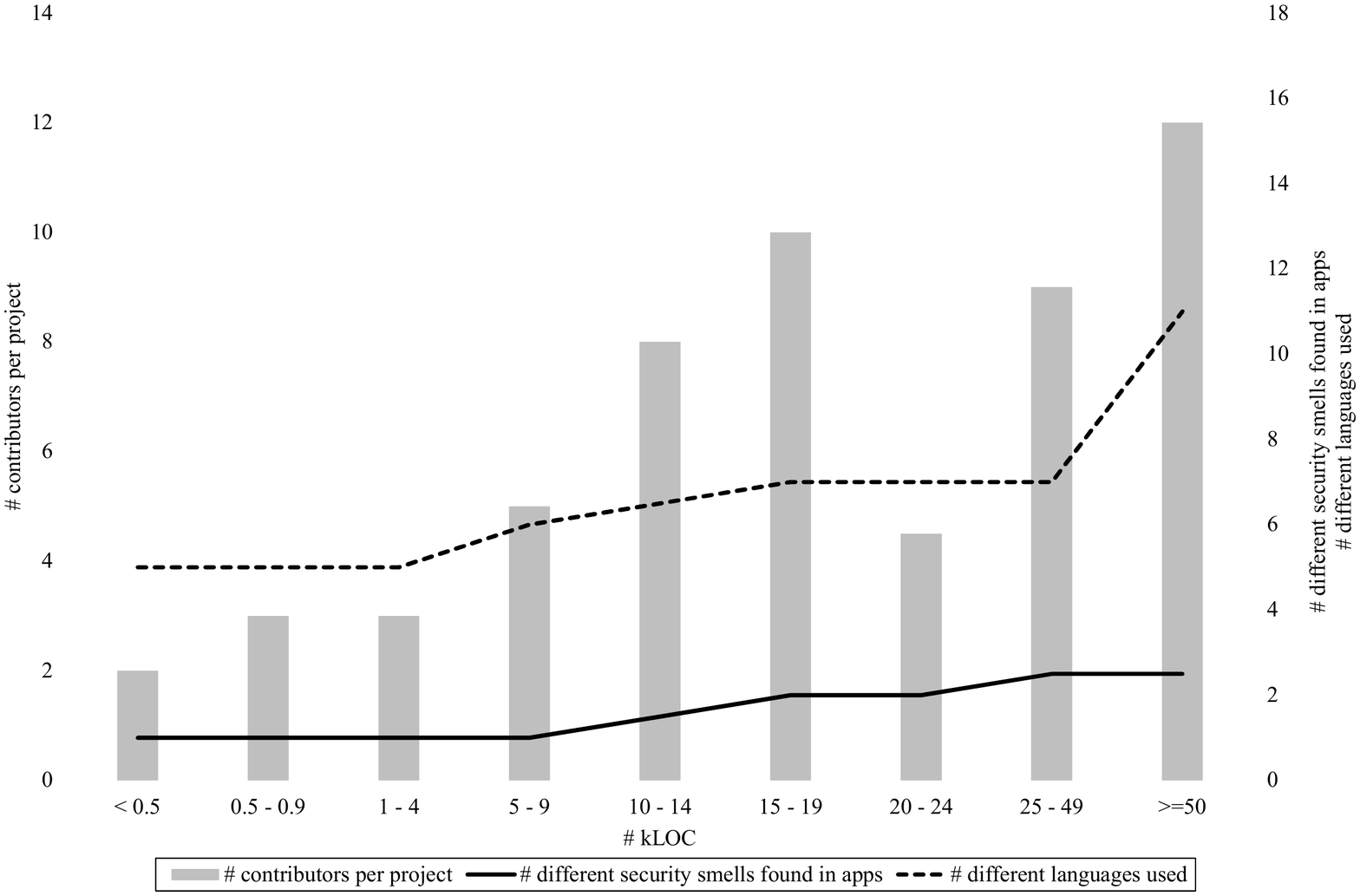}\\
\small (a) Relation of kLOC to contributors, ICC security smells, and used languages
\end{tabular}

\vspace{\floatsep}
\vspace{\floatsep}
\vspace{\floatsep}
\vspace{\floatsep}
\vspace{\floatsep}

\begin{tabular}{@{}c@{}}
% config for three plots on a page: width=8.25cm, trim = 2cm 2cm 2cm 2cm
\includegraphics[width=\columnwidth, trim = 2cm 2cm 2cm 2cm]{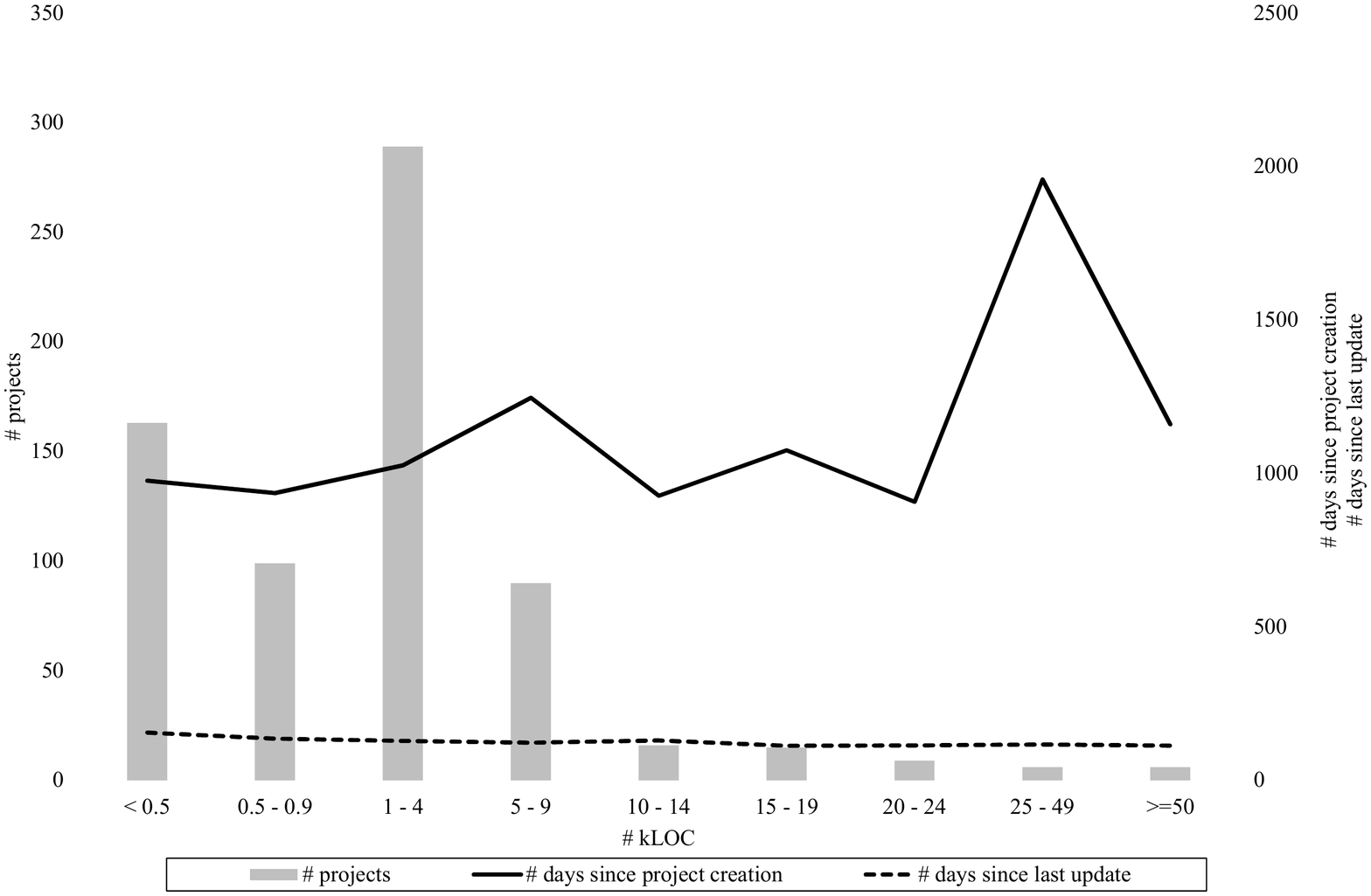}\\
\small (b) Relation of kLOC to number of projects, days since project creation, and days since last update
\end{tabular}

\caption{Different project properties in relation to kLOC}
\label{loc_size_correlations}
\end{figure}

Another popular indicator used in software analysis is the code size, which we measured in thousands of lines of code (kLOC) with the open-source tool cloc.\footnote{\url{https://github.com/AlDanial/cloc}}
As Android projects consist aside from source code of different configuration, resource and other utility files, we first ran the analysis of adopted software languages (\eg Java, Kotlin, XML) that required each of those items, before we excluded all elements except the Java code in the main Java source folders for the kLOC measurements.
We conjectured that we would see a trend of small teams developing small apps that are less likely to have problems.
In contrast, we expect that aging projects are more likely to have smells as they are larger than more recent ones.
\autoref{loc_size_correlations} illustrates the relation between the kLOC and other relevant properties.

In \autoref{loc_size_correlations}a we categorized projects according to their size on the x-axis, while the left y-axis displays contributors per project, and the right y-axis the number of different categories of security smells found in apps, and the number of different languages used.
We see a trend that larger projects rely on more contributors with a minor exception at 20-24 kLOC.
Furthermore, it is interesting to see that projects of up to 10 kLOC are maintained by five or fewer developers.
In addition, we see that larger projects tend to suffer from more smells, and those projects are also using more languages.
After a manual inspection of apps exploiting different languages we discovered that those apps are rather collections of frameworks, \eg for network penetration tests using a plethora of different tools written in different languages.

\autoref{loc_size_correlations}b uses the same feature for the x-axis, but presents on the left y-axis the number of projects, and on the right y-axis the number of days since project creation, and the number of days since last update.
The majority of projects in our dataset consist of less than 10 kLOC, and especially projects with 1-4 kLOC have been very prevalent, followed by apps that are less than 500 LOC.
Only six projects contained more than 50 kLOC.
Interestingly, we cannot derive clearly any major trend regarding the age of projects and LOC, although projects of 25-49 kLOC evidently are older than the others.
On the contrary, we can see a minor trend regarding the time since last update.
It appears that smaller apps are updated less frequently than larger apps.
We expect that the larger an application becomes, the more maintenance work is required due to library updates, obsolete external references, and content changes.

% --------------------------------------------------
\subsection{Manual Analysis}
\label{sec:manual-analysis}
To assess the performance of our tool and show how reliable these findings are to detect security vulnerabilities, we manually analyzed 100 apps.
We invited two participants to independently 
evaluate the precision and recall of our tool.
Participant A is a senior developer with more than 5 years of professional experience in development and security of mobile apps. 
Participant B is a junior developer with less than two years of experience in Java and C\# software development.
We provided both participants an introduction to Android security, and individually explained every smell in detail.
We subsequently selected the top 100 apps, that is more than 13\% of the whole corpus, with most smells in accordance with our ICC security smell list, for which we can say with 95\% confidence that the population's mean smell occurrences of the top 100 apps are between 3.04 and 3.48, while they are between 1.38 and 1.50 for the whole data set of about 732 apps.
Then we provided the participants with our tool, the sources of the top 100 apps, and a spreadsheet to record their observations.
Each participant was asked to import the sources of each app in Android Studio, which had been prepared to run a customized version of our analysis plug-in, to verify each reported smell according to the symptoms of any smell described in \autoref{sec:code-smells}.
We were also interested in vulnerability detection capability of security smells,\footnote{We define a vulnerability capability as the possibility a security issue can compromise a user's security and privacy.} thus the participants were asked to investigate if a security smell indicates the presence of a security vulnerability based on the vulnerability information available in the benchmarks.

% --------------------------------------------------
\subsubsection{Tool Evaluation}
While the assessment of true positives (TPs, reported code that is a smell) and false positives (FPs, reported code that is not a smell) requires participants to manually check only the tool's results, the extraction of true negatives (TNs, unreported code that is not a smell) and false negatives (FNs, unreported code that is a smell) is resource intensive and error prone.
Therefore to avoid an exhaustive code inspection, we developed a relaxed analysis that shows ICC-related APIs in the code to support the participants.

\begin{figure}
\includegraphics[width=\columnwidth, trim = 2cm 2cm 2cm 2cm]{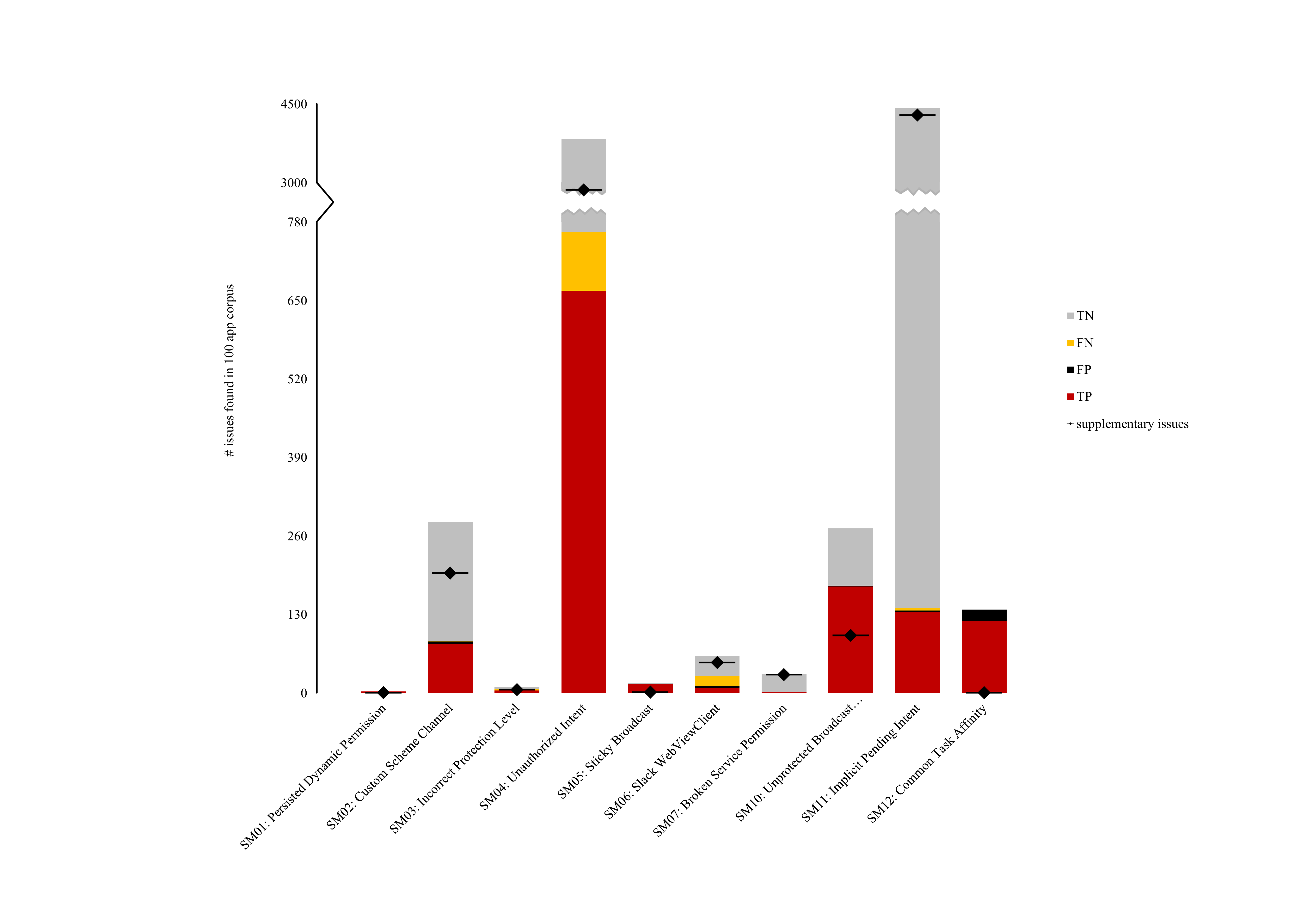}
\caption{Tool evaluation results}
\label{p_tool_performance}
\end{figure}

We obtained relatively high smell detection rates, especially for SM02, SM04, SM10, SM11 and SM12, as indicated by the TPs in \autoref{p_tool_performance}.
The reason is that these smells occur frequently and are straightforward to detect, mostly relying on some very specific method calls and permissions.

We encountered above average FPs in SM12 due to the intended use of task affinity features in apps that try to separate activities with empty task affinities.
This smell would require additional semantical, architectural, and UI information for proper assessment.
While some of the exposed activities are non-interactive, and thus supposedly secure, some of them are interactive and could be misused in combination with other spoofing techniques, like clickjacking, in which an adversary unexpectedly shows the exposed activity to trick users into providing unintended inputs.
In particular, call recorders and various client-server apps for chat, video streaming, home automation, and other network services have been affected by this issue.

Our participant had to check 7\,241 locations in the code to examine the TNs and FNs  in 100 apps. 
In more than 98.36\% of cases participants confirmed that there are no security smells beyond what the tool could identify; we consider this very low proportion of FNs, \ie 1.7\%, encouraging.

We are surprised to see only a few FNs in SM04 as we expected much more to appear due to the countless ways that intents can be created in Android.
A substantial number of FNs were missed because of complex chained executions and calls initiated from sophisticated UI related classes containing URIs.
For SM06, we discovered that the FNs have been frequently caused by lack of context, \eg unawareness of data sensitivity, or custom logic that does not mitigate the vulnerability.
For example, our tool was unable to verify the correctness of custom web page white-listing implementations for \code{WebView} browser components, which would actually reduce security if implemented incorrectly.

We did not encounter any instances of the two smells SM08 and SM09, that is, we retrieved zero reports on both of them for our 100 app dataset, hence, we excluded them for all subsequent plots and discussions in this subsection.

We could find common security smells while reviewing the feedback from the two participants, for example, that some apps were using \code{shouldOverrideUrlLoad\-ing} without URL white-listing to send implicit intents to open the device's default browser, rather than using their own web view for white-listed pages, thus fostering the risk of data leaks.
Another discovery was the use of regular broadcasts for intra-app communication.
For these scenarios, developers should solely rely on the \code{Local\-Broad\-cast\-Man\-ager} to prevent accidental data leaks.
The same applies for intents that are explicitly used for communication within the app, but do not include an explicit target, which would similarly mitigate the risk of data leaks.
Moreover, unused code represents a severe threat.
Several apps requested specific permissions without using them, increasing the impact of potential privilege escalation attacks.

% --------------------------------------------------
\subsubsection{Tool Performance}
\begin{figure}
\includegraphics[width=\columnwidth, trim = 2cm 2cm 2cm 2cm]{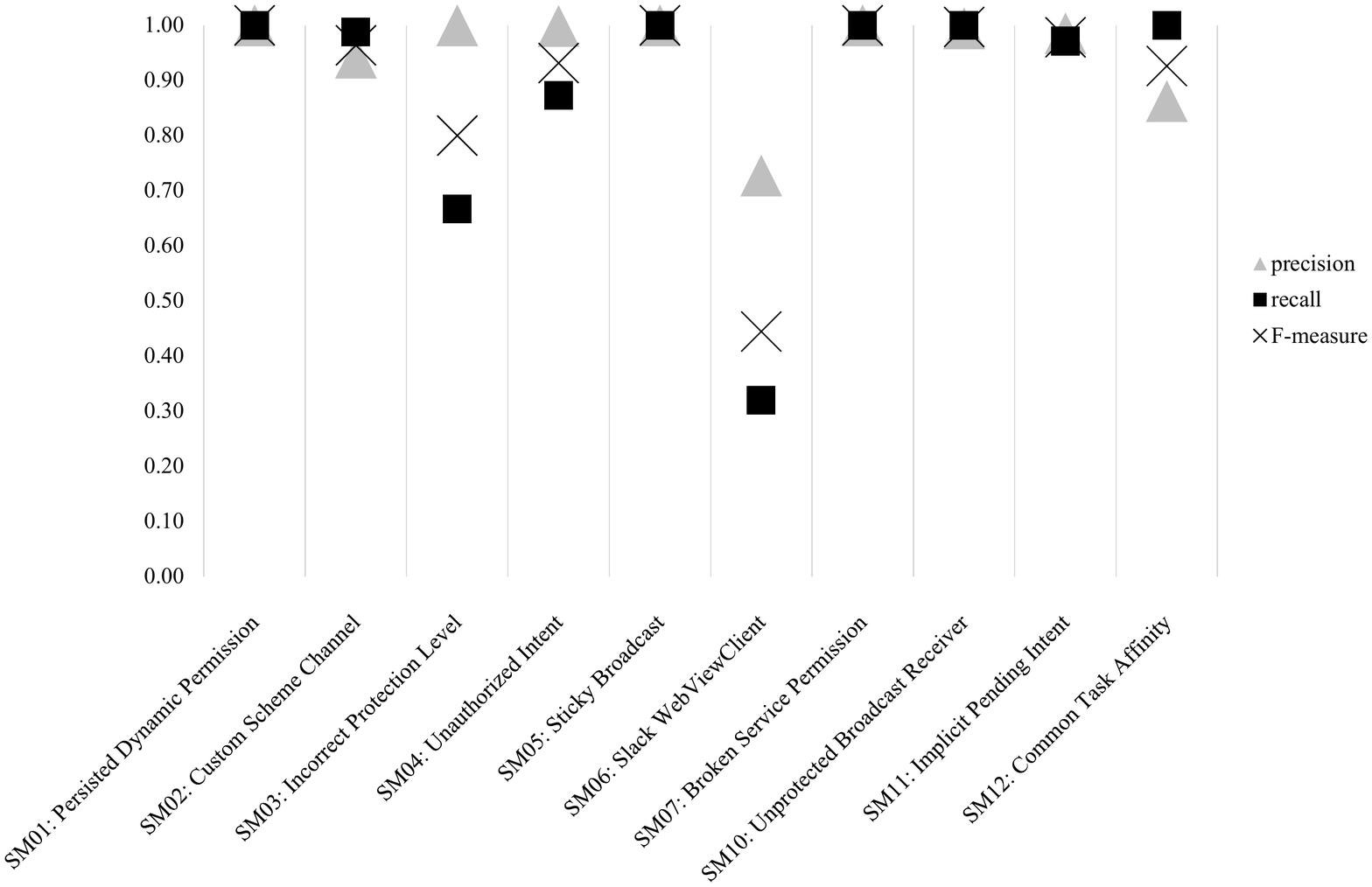}
\caption{Tool performance}
\label{p_tool_pr}
\end{figure}

\autoref{p_tool_pr} presents the tool's performance based on the precision, recall, and lastly the F-measure for existing smells in 100 apps.
All smells except SM03 and SM06 show outstanding results, nonetheless, some of them could be biased as a result of their low occurrences which is true for SM01 and SM07.
We performed a follow up manual investigation of SM03 and SM06.
Apparently, the detection of SM03 suffers from the difficulty to discern data sensitivity and the need to approximate the required protection level.
Besides that, SM06 is heavily affected by custom web API implementations that (mis)use security features, which are, in fact, not secure.

% --------------------------------------------------
\subsubsection{Smells and their Vulnerability Capability}
\label{SEC:vulnerability_capability}

\begin{figure}
\includegraphics[width=\columnwidth, trim = 2cm 1.5cm 2cm 1.5cm]{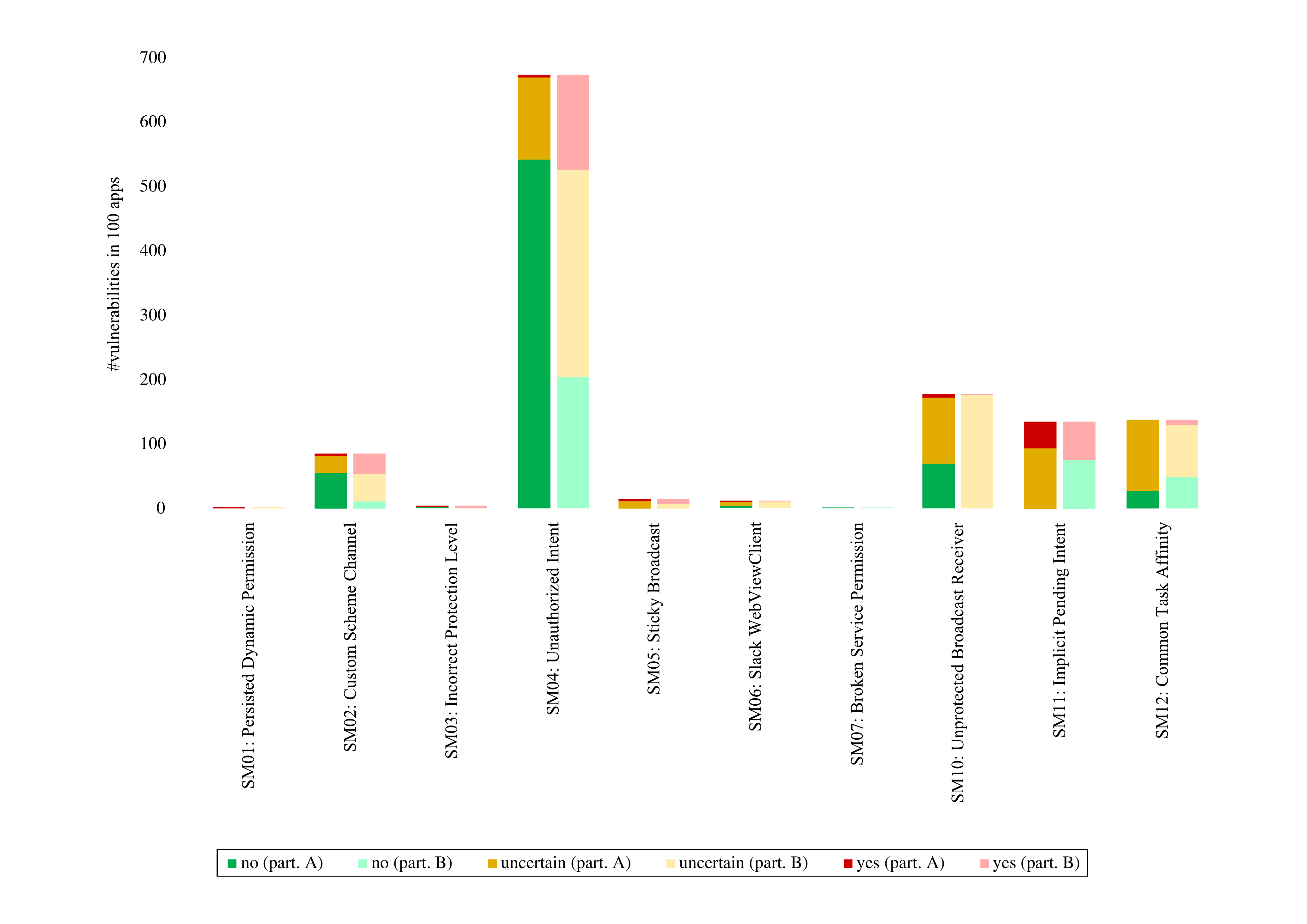}
\caption{Vulnerability capability of detected issues}
\label{p_tool_detected_vulnerabilities}
\end{figure}

\autoref{p_tool_detected_vulnerabilities} shows the vulnerability capability perception of both participants against the reported smells.
For each smell we show two grouped columns:
the left column reports the results from the more experienced participant A (PA), and the right column reports the results from the less experienced participant B (PB).
Each column consists of three different segments \emph{yes}, \emph{uncertain}, and \emph{no}.
The category \emph{yes} is used for all reported smells that introduce critical risks, such as plain-text exposure of user passwords through network sockets.
The \emph{uncertain} category is used for risks that potentially exist, and are challenging to inspect manually, for instance, vulnerabilities that require prerequisites for successful exploitation such as potentially dangerous user-defined schemes.
Finally, all smells assigned to the \emph{no} category are not vulnerable to any attacks, either because they do not contain any user information, or because they are sufficiently secure with respect to the participant's opinion.
Apps that send static non-sensitive information commonly match this category.
For all our considerations the participants were told to treat any user data as sensitive, since they could potentially contain sensitive information at run time.

According to the reports by PA, 38.5\% of smells represent potential threats, \ie \emph{uncertain} category, and only 5.3\% of smells represent critical threats, \ie \emph{yes} category. In other words, only about 44\% of security smells could lead to security vulnerabilities.

A further comparison of the reports between the two participants shows that they expect somewhat similar risks for the smell categories SM05, SM07, and SM12, whereas the participants tended to interpret diversely the threat caused by \emph{Custom Scheme Channel}, \emph{Unauthorized Intent}, and \emph{Slack WebViewClient} smells.
We reviewed the feedback of the participants and discovered that for \emph{Custom Scheme Channel} predefined system schemes are considered less harmful for PA (category \emph{no}), while PB assigns them to the category \emph{uncertain}.
For \emph{Unauthorized Intent} PB assessed the risks similar to PA, however, PB encountered difficulties to predict adequately the threat capability of many intent instances, thus PB assigned them to section \emph{uncertain}.
For the smell \emph{Slack WebViewClient} PB per\-form\-ed a conservative risk assessment by not assigning any custom security feature implementations to \emph{no}, instead PB assigned them to \emph{uncertain}, unlike PA who conc\-lud\-ed many of them as secure.
An example thereof is an app with a network security penetration test suite that requires opening insecure web pages for security validation purposes.

It is interesting to observe that PB, in contrast to PA, does consider fewer instances as harmful for SM11 and SM12.
For the first smell \emph{SM11: Implicit Pending Intent} PB considers intents with assigned actions frequently as secure, while PA considers them as potential risk, which is more accurate.
For the second smell \emph{SM12: Common Task Affinity} PB considers most apps that used empty task affinity properties as secure, while PA performed a more thorough analysis of the UI and considered additionally the misuse capability of such exposed views, which resulted in many assignments to the category \emph{uncertain}.
We conclude that the very complex and flexible ICC implementation provided by Android overwhelms inexperienced developers, even worse, it could mislead those developers to create insecure code due to their misunderstanding.

Overall, most of the vulnerabilities seem to emerge from SM10 and SM11 that collectively contribute to more than 72\% of all detected critical issues.
On the other hand, SM04 on its own provides with 77\% the largest proportion of false alarms regarding vulnerability capabilities.

% --------------------------------------------------
\subsection{Threats to Validity}

One important threat to validity is the completeness of this study, \ie whether we could identify and study all related papers in the literature.
Although we could not review all publications, we strived to explore top-tier software engineering and security journals and conferences as well as highly-cited work in the field.
For each relevant paper we also recursively looked at both citing and cited papers.
Moreover, to ensure that we did not miss any important paper, for each identified issue we further constructed more specific queries and looked for any new paper on GoogleScholar.

We were only interested in studying benign apps as in malicious ones it is unlikely that developers will spend any effort to accommodate security concerns.
Thus, we merely collected apps that were available on GitHub and the F-Droid repository.
However, our dataset may still have malicious apps that evaded the security checks of the community or the marketplace.

We analyzed the existence of security smells in the source code of an app, whereas third-party libraries could also introduce smells.

Our analysis is intra-procedural and suffers from inherent limitations of static analysis.
Moreover, many security smells actually constitute security risks only if they deal with sensitive data, but our analysis cannot determine such sensitivity.

The Android Lint tool we used for the analysis is prone to errors that could lead to FNs, for example, when Android Lint crashes due to file parsing issues, an immediate termination of the inspection occurs which could cause some misses.

Finally, the fact that the results of our analysis tool are validated against manual analysis performed by the authors is a threat to construct validity through potential bias in experimenter expectancy.
We mitigated this threat by including an external participant in the process in addition to the co-author who simultaneously played the senior developer's role.

% ==================================================
\section{Related Work} \label{sec:related-work}

Reaves~\etal studied Android-specific challenges to program analysis, and assessed existing Android application analysis tools.
They found that these tools mainly suffer from lack of maintenance, and are often unable to produce functional output for applications with known vulnerabilities~\cite{Reaves:2016}.
Li~\etal studied the state-of-the-art work that statically analyses Android apps~\cite{Li:2017b}.
They found that much of this work supports detection of private data leaks and vulnerabilities, a moderate amount of research is dedicated to permission checking, and only three studies deal with cryptography issues.
Unfortunately, much state-of-the-art work does not publicly share the concerned artifacts.
Linares-Vasquez~\etal mine 660 Android vulnerabilities available in the official Android bulletins and their CVE details,\footnote{\url{http://cve.mitre.org} --- Common Vulnerabilities and Exposures, a public list of known cyber-security vulnerabilities.
} and present a taxonomy of the types of vulnerabilities~\cite{Linares-Vasquez:2017}.
They report on the presence of those vulnerabilities affecting the Android OS, and acknowledge that most of them can be avoided by relying on secure coding practices.
Finally, Sadeghi~\etal review 300 research papers related to Android security, and provide a taxonomy to classify and characterize the state-of-the-art research in this area~\cite{Sadeghi:2016}.
They find that 26\% of existing research is dedicated to vulnerability detection, but each study is usually concerned with specific types of security vulnerabilities.
Our work expands on such studies to provide practitioners with an overview of the security issues that are inherent in insecure programming choices.

Some research is devoted to educating developers in secure programming.
Xie~\etal interviewed 15 professional developers about their software security knowledge, and realized that many of them have reasonable knowledge but do not apply it as they believe it is not their responsibility~\cite{Xie:2011}.
Weir~\etal conducted open-ended interviews with a dozen app security experts, and determined that app developers should learn analysis, communication, dialectics, feedback, and upgrading in the context of security~\cite{Weir:2016b}.
Witschey~\etal surveyed developers about their reasons for adopting or not adopting security tools~\cite{Witschey:2015}.
Interestingly, they found the perceived prestige of security tool users and the frequency of interaction with security experts to be important for promoting security tool adoption.
Acar~\etal suggest a high-level research agenda to achieve usable security for developers.
They propose several research questions to elicit developers' attitudes, needs and priorities in the area of security~\cite{Acar:2016b}.
Our work is complementary to these studies in the sense that we provide an initial assessment of developers' security knowledge, and we highlight the significant role of developers in making apps more secure.
	
Numerous researchers have dedicated their work to detecting common ICC vulnerabilities.
Despite the fact that their expression has changed over time, the vulnerability classes have remained largely the same.
Chin \etal discuss the ICC implementation of Android and examine closely the interaction between sent and received ICC messages~\cite{Chin:2011}.
Despite the fact that their work is based on a small corpus containing only 20 apps, they were able to detect various denial-of-service issues in numerous application components, and conclude that the message-passing system in Android enables rich applications, and encourages component reuse, while leaving a large potential for misuse when developers do not take any precautions.

Felt \etal discovered that permission re-delegation, also known as confused deputy or privilege escalation attack, is a common threat, and they pose OS level mitigations conceptually similar to the same origin policy in web browsers~\cite{Felt:2011}.
The community aimed on the one hand for preciseness, as countless tools to detect these flaws in ICC have been released, notably Epicc~\cite{Octeau:2013} and IccTA~\cite{Li:2015} with a significantly improved precision.
On the other hand, the app coverage began to play a major role, as in the work of Bosu \etal who recently discovered with their tool inadequate security measures, including privilege escalation vulnerabilities, among inter-app data-flows from 110\,000 real-world apps~\cite{Bosu:2017}.

Along with passive analysis, active countermeasures and attacks have emerged in the scientific community.
Garcia \etal crafted a state-of-the-art tool to automatically detect and exploit vulnerable ICC interfaces to provoke denial-of-service attacks amongst others~\cite{Garcia:2017}.
They identified exploits for more than 21\% of all apps appraised as vulnerable.
Xie \etal presented a bytecode patching framework that incorporates additional self-contained permission checks avoiding privilege issues during runtime, generating a remarkably low computational overhead~\cite{Xie:2017}.
Ren \etal successfully investigated design glitches in the multitasking implementation of Android, uncovering task hijacking attacks that affected every OS release and were potentially duping user perception~\cite{Ren:2015}.
They considered in particular the taskAffinity and taskParentReparenting attributes of the manifest file that allow views to be dynamically overlaid on other apps, and provided proof-of-concept attacks.
Wang \etal assessed the threat of data leakage on Apple iOS and Android mobile platforms and show serious attacks facilitated by the lack of origin-based protection on ICC channels~\cite{Wang:2013b}.
Interestingly, they found effective attacks against apps from such major publishers as Facebook and Dropbox, and more importantly, indicate the existence of cross-platform ICC threats.
Researchers have found interest in reinforcing the Android ICC core framework.
Khadiranaikar \etal propose a certificate-based intent system relying on key stores that guarantee integrity during message exchanges~\cite{Khadiranaikar:2017}.
In addition to securing the ICC-based communication, Shekhar \etal proposed a separation of concerns to reduce the susceptibility for manipulation of Android apps, by explicitly restricting advertising frameworks~\cite{Shekhar:2012}.
Ahmad \etal elaborated on problematic ICC design decisions on Android, and found that missing consistent message types and conformance checking, unpredictable message interactions, and a lack of coherent versioning could break inter-app communication and pose a severe risk~\cite{Ahmad:2016}.
They recommend a centralized message-type repository that immediately provides feedback to developers through the IDE.

In summary, existing studies have often dealt with a specific issue, whereas we cover a broader range of issues, making the results more actionable for practitioners.
Moreover, previous work often overwhelms developers with many identified issues at once, whereas we provide feedback during app development where developers have the relevant context.
Such feedback makes it easier to react to issues, and helps developers to learn from their mistakes~\cite{Tymchuk:2018}.

% ==================================================
\section{Conclusion} \label{sec:conclusion}

We have reviewed ICC security code smells that threaten Android apps, and implemented a linting plug-in for Android Studio that spots such smells, by linting affected code parts, and providing just-in-time feedback about the presence of security code smells.

We applied our analysis to a corpus of more than 700 open-source apps.
We observed that only small teams are capable of consistently building software resistant to most security code smells, and fewer than 10\% of apps suffer from more than two ICC security smells.
We discovered that updates rarely have any impact on ICC security, however, in case they do, they often correspond to new app features.
Thus developers have to be very careful about integration of new functionality into their apps.
Moreover, we found that long-lived projects suffer from more issues than recently created ones, except for apps that are updated frequently, for which that effect is reversed.
We advise developers of long-lived projects to continuously update their IDEs, as old IDEs have only limited support for security issue reports, and therefore countless security issues could be missed.

A manual investigation of 100 apps shows that our tool successfully finds many different ICC security code smells, and about 43.8\% of them in fact represent vulnerabilities, thus it constitutes a reasonable measure to improve the overall development efficiency and software quality.

We recommend security aspects such as secure default values and permission systems, to be considered in the initial design of a new API, since this would effectively mitigate many issues like the very prevalent Common Task Affinity smell.
We plan to explore the extent to which APIs can be made secure by design.
While we analyzed the existence of ICC security smells in apps, studying their absence, \ie secure ICC uses, could offer different insights that we plan to pursue in future.
Moreover, we are interested in evaluating the usefulness of our tool during a security audit process, as well as in an app development session.

\begin{acknowledgements}
We gratefully acknowledge the financial support of the Swiss National Science Foundation for the project ``Agile Software Analysis'' (SNSF project No.\,200020-162352, Jan 1, 2016 - Dec.\,30, 2018).
We also thank Astrid Ytrehorn for her contribution to the empirical study. 
\end{acknowledgements}

\bibliographystyle{plain}
\bibliography{reference} 

\end{document}